\documentclass[aps,pre,twocolumn,superscriptaddress,citesort]{revtex4-2}
\usepackage{graphicx}
\usepackage{amsmath}

\usepackage{bm}

\usepackage{hyperref}
\usepackage{xcolor}
\definecolor{tab_blue}{HTML}{1F77B4}
\hypersetup{%
  breaklinks=true,
  colorlinks=true,
  linkcolor=tab_blue,
  filecolor=tab_blue,
  urlcolor=tab_blue,
  citecolor=tab_blue,
}

\begin{document}

\title{Efficient point-based simulation of four-way coupled particles in turbulence at high number density}
\author{Xander M. de Wit}
\affiliation{Fluids and Flows group and J.M. Burgers Center for Fluid Mechanics, Department of Applied Physics and Science Education, Eindhoven University of Technology, 5600 MB Eindhoven, Netherlands}
\author{Rudie P. J. Kunnen}
\affiliation{Fluids and Flows group and J.M. Burgers Center for Fluid Mechanics, Department of Applied Physics and Science Education, Eindhoven University of Technology, 5600 MB Eindhoven, Netherlands}
\author{Herman J. H. Clercx}
\affiliation{Fluids and Flows group and J.M. Burgers Center for Fluid Mechanics, Department of Applied Physics and Science Education, Eindhoven University of Technology, 5600 MB Eindhoven, Netherlands}
\author{Federico Toschi}
\email{f.toschi@tue.nl}
\affiliation{Fluids and Flows group and J.M. Burgers Center for Fluid Mechanics, Department of Applied Physics and Science Education, Eindhoven University of Technology, 5600 MB Eindhoven, Netherlands}
\affiliation{CNR-IAC, I-00185 Rome, Italy}
\date{\today}

\begin{abstract}
In many natural and industrial applications, turbulent flows encompass some form of dispersed particles. Although this type of multiphase turbulent flow is omnipresent, its numerical modeling has proven to be a remarkably challenging problem. Models that fully resolve the particle phase are computationally very expensive, strongly limiting the number of particles that can be considered in practice. This warrants the need for efficient reduced order modeling of the complex system of particles in turbulence that can handle high number densities of particles. Here, we present an efficient method for point-based simulation of particles in turbulence that are four-way coupled. In contrast with traditional one-way coupled simulations, where only the effect of the fluid phase on the particle phase is modeled, this method additionally captures the back-reaction of the particle phase on the fluid phase, as well as the interactions between particles themselves. We focus on the most challenging case of very light particles or bubbles, which show strong clustering in the high-vorticity regions of the fluid. This strong clustering poses numerical difficulties which are systematically treated in our work. Our method is valid in the limit of small particles with respect to the Kolmogorov scales of the flow and is able to handle very large number densities of particles. This methods paves the way for comprehensive studies of the collective effect of small particles in fluid turbulence for a multitude of applications.
\end{abstract}
\maketitle	

\section{Introduction}

Understanding the behavior of particles in turbulent flows is paramount in many fields, including engineering, meteorology, oceanography, and astrophysics. The delicate interplay between particle dynamics and the background turbulence encompasses a rich phenomenology \cite{Balachandar2010,Toschi2009,Mathai2020,Brandt2022,Benzi2023}. To accurately describe this complex system, in general, one needs to consider the reciprocal interactions between fluid and particles, as well as the mutual interactions between particles. From the computational point of view, modeling this complete interplay turns out to be challenging and often computationally prohibitive.

Therefore, most numerical studies historically have focused on the so-called one-way coupling paradigm, where only the effect of turbulence on the particle dynamics is modeled but not vice-versa. This is a reasonable approximation for small particles in the very dilute limit and was shown to recover much of the phenomenology observed in experiments, such as preferential concentration, particle dispersion and clustering \cite{Maxey1987,Crisanti1992,Balkovsky2001,Calzavarini2008}. In some cases, however, one may be particularly interested in the back-reaction that the particle phase has on the underlying fluid phase. This requires a so-called two-way coupled simulation, where also this back-reaction is explicitly modeled. The most accurate way to model this back-reaction is by fully resolving the geometry of the particles and enforcing the no-slip condition at the particle-fluid interface, e.g. via approaches such as the Immersed Boundary Method \cite{Peskin1972,Fadlun2000,Orlandi2006,Seo2011,Verzicco2023}. Though possibly being very accurate, this method is also computationally very expensive and in practice, this limits the number of particles that can be considered to around $\mathcal{O}(10-1000)$ \cite{Verzicco2023}. To study the collective effect of a large number of small particles dispersed in the turbulent flow, this is insufficient, warranting the need for more computationally efficient reduced order modeling.

In this work, we therefore consider instead a point-particle based model which includes the back-reaction force on the fluid and that can easily be extended to millions of particles \cite{Eaton2009,VandenBerg2009,Monchaux2017}. To accurately handle substantial volume fractions of particles and/or to handle cases of strong particle clustering, however, we need to go one step further and model also the interactions between particles, bringing us in the realm of the so-called four-way coupled simulations.

We perform Direct Numerical Simulations (DNS) of the fluid phase with spatio-temporal velocity field $\bm{u}(\bm{x},t)$ using a standard pseudospectral method that solves the incompressible Navier-Stokes equations
\begin{gather} \label{eq:NS}
    \frac{\mathrm{D} \bm{u}}{\mathrm{D} t} = -\bm{\nabla} p + \nu \Delta \bm{u} + \bm{f} + \bm{f}_p,\\
    \bm{\nabla}\cdot\bm{u}=0,
\end{gather}
with $\mathrm{D} \bm{u}/ \mathrm{D} t \equiv \partial \bm{u} / \partial t + \bm{u} \cdot \bm{\nabla} \bm{u}$ denoting the material derivative. Here, $p$ is the local pressure, $\nu$ denotes the kinematic viscosity of the fluid and $\bm{f}$ is the forcing of the fluid, which we choose to be \mbox{$\hat{\bm{f}}(\bm{k}) = \epsilon \hat{\bm{u}}(\bm{k}) / \sum_{k_f \leq |\bm{k}| < k_f+1} |\hat{\bm{u}}(\bm{k})|^2$}, acting on the low wavenumbers $k_f \leq |\bm{k}| < k_f+1$, ensuring a constant energy injection rate $\epsilon$ \cite{Lamorgese2005}. The feedback force of the particle phase on the fluid phase is captured in $\bm{f}_p$, which is derived in Sec.~\ref{sec:delta_coupling}. For time integration, we use a 2\textsuperscript{nd} order Adam-Bashfort scheme, while the viscous term is integrated exactly using integrating factors. We resolve the turbulence well down to the Kolmogorov scale $\eta$ by ensuring that for the highest resolved wavenumber we have $k_\textrm{max}\eta \approx 3$, such that the grid spacing coincides with the Kolmogorov scale. For details on the Eulerian scheme, see Appendix~\ref{sec:eulerian_solver}.

To describe the particle phase, we resort to the Maxey-Riley equations of motion for dispersed particles \cite{Maxey1983}. This point-based model requires that the particle has a sufficiently small size to see a smooth velocity field, which in practice amounts to an $\mathcal{O}(1)$ multiple of the Kolmogorov scale $\eta$. We consider the limit where we can neglect gravity and for simplicity we also disregard the Basset history force that corrects for the unsteady boundary layer around the particle \cite{Maxey1983, vanHinsberg2017}. We then retain only the Stokes drag term and the combined pressure gradient and added mass term, yielding
\begin{align}\label{eq:particle}
    \frac{\mathrm{d} \bm{V}}{\mathrm{d} t} &= \beta \frac{\mathrm{D} \bm{u}(\bm{X},t)}{\mathrm{D} t} -\frac{1}{\tau_p}(\bm{V}-\bm{u}(\bm{X},t)),\\
    \frac{\mathrm{d} \bm{X}}{\mathrm{d} t} &= \bm{V} + \textrm{(collisions)}.
\end{align}
Here $\bm{V}$ and $\bm{X}$ denote the velocity and position of the particle, respectively. Furthermore, the density ratio between the particle and the fluid is parameterized by $\beta = 3/(1+2\rho_p / \rho_f)$ with $\rho_p$ and $\rho_f$ the density of the particle and fluid phase, respectively. Finally, $\tau_p$ denotes the particle response time, which for spheres is given by $\tau_p = D^2/(12 \beta \nu)$ with $D$ the particle diameter. The latter is typically non-dimensionalized with respect to the Kolmogorov timescale of the fluid $\tau_\eta$, yielding the Stokes number $\textrm{St}=\tau_p/\tau_\eta$. The fluid velocity and its material derivative is evaluated at the particle position using spline interpolation \cite{vanHinsberg2013}. Note that in the integration of the particle position, we also take into account the particle-particle collisions, which is treated in Sec.~\ref{sec:excluded_volume}, completing our four-way coupling method.

Arguably among the most striking phenomena of the behavior of particles in turbulence is the preferential concentration of particles that are heavier or lighter than the fluid, observed at moderate $\textrm{St}\sim\mathcal{O}(1)$. While heavy particles ($0 \leq \beta < 1$) tend to preferentially concentrate in high-straining regions, light particles ($1 < \beta \leq 3$) show the opposite behavior, strongly concentrating in high-vorticity regions of the turbulent flow \cite{Calzavarini2008}. While the method presented here is valid for small particles with any density ratio and/or $\textrm{St}$, we focus on bubbles $\beta=3$ at moderate $\textrm{St}=1$ \footnote{Note that while physically, assuming Stokes drag on a sphere, the particle size $D/\eta$ becomes fixed as soon as one assumes a certain $\textrm{St}$ and $\beta$, but numerically we can independently vary the particle size used in the two-way and four-way coupling to study its influence and to ensure that the particle remains sufficiently small.}, which show the strongest clustering due to their preferential concentration in the vortex filaments of the turbulent flow \cite{Aliseda2011,Toschi2009}. Understanding the dynamics of bubbles in turbulence is relevant for many industrial and natural processes and it is widely studied experimentally and numerically \cite{Mercado2012, Chouippe2014, Mathai2016, Loisy2017, Mathai2018}. As we shall show, the strong clustering poses additional difficulties to the four-way coupling scheme and thus acts as a \textit{worst-case-scenario} benchmark to our method.

Sec. \ref{sec:part-fluid} treats the implementation of the particle-fluid interaction, while Sec. \ref{sec:part-part} details the scheme for the particle-particle interaction. Finally, Sec. \ref{sec:example} covers an example the modulation of the turbulent energy spectrum under the influence of four-way coupled bubbles and conclusions are drown in Sec. \ref{sec:conclusion}.

\section{Particle-fluid interaction}\label{sec:part-fluid}

\subsection{Momentum conservation: the delta-coupling} \label{sec:delta_coupling}

\begin{figure*}
    \centering
    \includegraphics[width=0.8\linewidth]{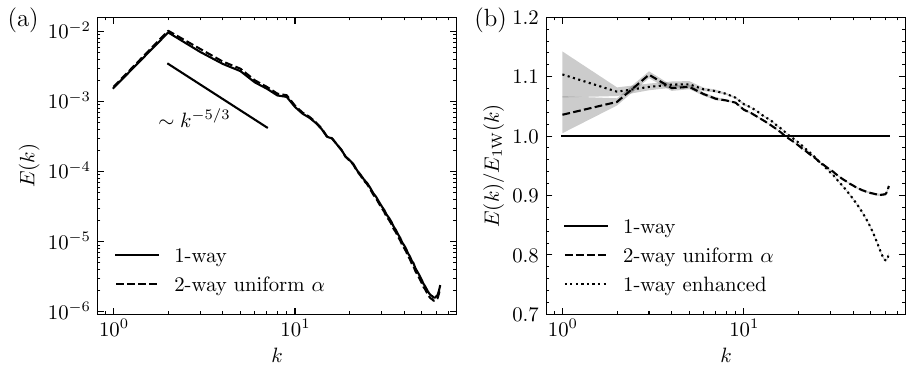}
    \caption{Panel (a): Energy spectra of the one-phase turbulent flow and the two-way coupled bubbly turbulent flow with uniform volume fraction of bubbles $\alpha_c=0.1$. Panel (b): the two-way coupled energy spectrum compensated by the one-way energy spectrum reveals that the two-way coupling enhances the energy spectrum in the forcing and inertial range, while it attenuates the spectrum in the dissipative range. It furthermore shows that the two-way coupled system is in close agreement with a one-way system where the energy injection rate $\epsilon$ and viscosity $\nu$ are enhanced by a factor $1/(1-\alpha_c)$, consistent with the theoretical prediction Eq.~\eqref{eq:two-way_enhanced}. The resolution is $N^3=128^3$ and the number of particles is $N_p=2\,097\,152\,(=128^3)$. The shaded regions denote statistical errorbars.}
    \label{fig:spectra_2w_bench}
\end{figure*}

The fluid phase is resolved on an Eulerian grid using a pseudospectral method as laid out in Appendix~\ref{sec:eulerian_solver}. To account for the back-reaction of the particle on the fluid, we need to consider the particle as a source of momentum. The particle phase then poses an additional term in the Navier-Stokes equations that follows from the conservation of momentum between the particle and fluid phase. The open question that remains is how the total momentum that is transferred from the particle to the fluid is distributed in space. In general, this depends on the size and shape of the particle and requires resolving the full particle as is done e.g. in the Immersed Boundary Method. In the limit where the particle is small with respect to the Kolmogorov length of the underlying turbulent flow, however, the spatial distribution of the back-coupling force conveniently reduces to a $\delta$-distribution \cite{Boivin1998,Eaton2009}. This can be treated efficiently in the DNS.

To derive the expression for the two-way coupling, we start from the integral conservation of momentum in the fluid phase with volume $\mathcal{V}_f$ with one submerged particle with volume $\mathcal{V}_p$ and surface $S_p$ \cite{Mazzitelli2003}
\begin{equation} \label{eq:momentum_fluid}
    \int_{\mathcal{V}_f} \rho_f \frac{\mathrm{D} \bm{u}}{\mathrm{D} t} \mathrm{d} \mathcal{V}=\int_{\mathcal{V}_f} \rho_f \left[ -\bm{\nabla} p + \nu \Delta \bm{u} + \bm{f} \right] \mathrm{d} \mathcal{V} +\int_{S_p} \boldsymbol{\sigma} \cdot \bm{n} \mathrm{d} S,
\end{equation}
with $\bm{n}$ the surface normal of the particle and $\bm{\sigma}$ the fluid stress. Then, for the particle phase, we can write momentum conservation as
\begin{equation} \label{eq:momentum_particle}
    \rho_p \mathcal{V}_p \frac{\mathrm{d} \bm{V}}{\mathrm{d} t}=-\int_{S_p} \boldsymbol{\sigma} \cdot \bm{n} \mathrm{d} S.
\end{equation}
When evaluated for the particle, this results in the Maxey-Riley equations of motion Eq.~\eqref{eq:particle}.
For the fluid phase, however, in general, substituting Eq.~\eqref{eq:momentum_particle} into Eq.~\eqref{eq:momentum_fluid} and extending the integral over the full domain $\mathcal{V}=\mathcal{V}_p+\mathcal{V}_f$ gives
\begin{eqnarray} \label{eq:momentum_full}
    \int_{\mathcal{V}} \rho_f &&\frac{\mathrm{D} \bm{u}}{\mathrm{D} t} \mathrm{d} \mathcal{V} =\int_{\mathcal{V}} \rho_f \left[ -\bm{\nabla} p + \nu \Delta \bm{u} + \bm{f} \right] \mathrm{d} \mathcal{V} \nonumber \\ &&+\int_{\mathcal{V}} \left[ \rho_f \frac{\mathrm{D} \bm{u}}{\mathrm{D} t} - \rho_p \frac{\mathrm{d} \bm{V}}{\mathrm{d} t} \right] \mathcal{V}_p \delta\left(\bm{x} - \bm{X}(t) \right) \mathrm{d} \mathcal{V}.
\end{eqnarray}

The total two-way coupling force is then given by a sum over all particles $i$ \cite{Maxey1994,Mazzitelli2003}
\begin{equation} \label{eq:bubble_force}
    \bm{f}_p = \sum_i^{N_p} \left[ \frac{\mathrm{D} \bm{u}}{\mathrm{D} t} - \frac{\rho_p}{\rho_f} \frac{\mathrm{d} \bm{V}_i}{\mathrm{d} t} \right] \mathcal{V}_p \delta\left(\bm{x} - \bm{X}_i(t) \right).
\end{equation}
Note that the two-way coupling term not only reflects the momentum exchange (second term in the brackets), but also has an added mass term (first term in the brackets).

The two-way coupling force Eq.~\eqref{eq:bubble_force} can be computed by evaluating the Lagrangian particle acceleration and the Eulerian material derivative, interpolated at the position of the particle. To evaluate the $\delta$-function, practically, it is smoothed by extrapolating to the eight nearest grid cells around the particle center using linear volume-weighting, summing to $1/\mathcal{V}_c$ with $\mathcal{V}_c$ the volume of one Eulerian computational cell \cite{Eaton2009,VandenBerg2009,Monchaux2017}. This ensures that the $\delta$-function integrates to unity. By summing over all particles, we can then obtain $\bm{f}_p$ evaluated on the full Eulerian grid, after taking into account the communication at the boundaries of the computational processes. As a final step, $\bm{f}_p$ is projected onto its divergence-free part to retain incompressibility. Semantically, this entails a redefinition of the pressure that compensates the divergent part.

An important issue for two-way coupled point-based models that should be pointed out is the problem of self-induction \cite{Boivin1998, Eaton2009, Tom2022}. The Maxey-Riley equations Eq.~\eqref{eq:particle} are defined with respect to the flow field that is perturbed by all other particles except for the particle under consideration itself. However, in practice, it is impossible to disentangle the contributions from the different disturbance fields created by each particle, so we have to make the approximation that the self-induction is negligible. In Appendix~\ref{sec:self_ind}, we justify this assumption by following the test that is laid out in Ref.~\cite{Mazzitelli2003}, by comparing the diffusion of particles that are back-coupled to the fluid and particles that are not back-coupled simultaneously in the same simulation.

\subsection{Two-way coupling validation: homogeneous distribution of bubbles}
As a point of validation, we can consider a homogeneous distribution of bubbles. For bubbles $\rho_p / \rho_f=0$, so that only the added mass term in Eq.~\eqref{eq:bubble_force} remains. Note that in this case, the particle equation of motion vanishes from the back-reaction $\bm{f}_p$ and only the spatial distribution of the bubbles becomes important. Indeed, in a continuum description, we obtain
\begin{equation}
    \bm{f}_p = \alpha(\bm{x},t) \frac{\mathrm{D} \bm{u}}{\mathrm{D} t},
\end{equation}
with $\alpha(\bm{x},t) \equiv \sum_i^{N_p} \mathcal{V}_p \delta\left(\bm{x} - \bm{X}_i(t) \right)$ the local volume fraction of bubbles. This yields
\begin{equation}\label{eq:two-way_bubs}
    \frac{\mathrm{D} \bm{u}}{\mathrm{D} t} = - \bm{\nabla} p + \nu \Delta \bm{u} + \bm{f} + \alpha(\bm{x},t) \frac{\mathrm{D} \bm{u}}{\mathrm{D} t}.
\end{equation}
We can then analytically compute the effect of the two-way forcing on the underlying fluid if we enforce a homogeneous distribution of particles $\alpha(\bm{x},t) = \alpha_c$. Then Eq.~\eqref{eq:two-way_bubs} reduces to
\begin{equation}\label{eq:two-way_enhanced}
    \frac{\mathrm{D} \bm{u}}{\mathrm{D} t} = - \frac{1}{1-\alpha_c} \bm{\nabla} p + \frac{1}{1-\alpha_c} \nu \Delta \bm{u} + \frac{1}{1-\alpha_c} \bm{f}.
\end{equation}
That is, solving the two-way coupled system should become equivalent to solving a one-phase system, where both the viscosity $\nu$ as well as the energy injection rate $\epsilon$ are enhanced by a factor $1/(1-\alpha_c)$. The prefactor to the pressure gradient can be absorbed in a redefinition of the effective pressure and thus has no dynamical influence.

This correspondence is confirmed numerically as laid out in Fig.~\ref{fig:spectra_2w_bench} by considering the turbulent kinetic energy spectrum, validating the implementation of our two-way coupling algorithm for the case where we enforce a homogeneous distribution of immobilized bubbles. Only in the highest wavenumber range, some quantitative differences between the two-way coupled system and the enhanced one-way coupled system can be observed. We attribute these discrepancies to discretization errors, since the two-way coupling involves interpolation and discrete time-stepping, while the viscous term is integrated exactly using integrating factors. Indeed, we confirmed that upon refining the spatiotemporal resolution, this remaining discrepancy at high wavenumbers vanishes (not shown).

We thus understand that the added mass effect of bubbles enhances the turbulence in the forcing and inertial range, while it simultaneously also enhances the viscous dissipation, attenuating the turbulence in the dissipative range. However, while we have shown this for the case of a uniform density of bubbles, in reality, due to preferential sampling of bubbles in vortex filaments, we should expect the back-reaction of bubbles to be highly localized in space, which can change the global influence of the bubbles on the underlying turbulence.

From the numerical point of view, this poses two additional problems in the pursuit to obtain reliable simulations of bubbly turbulence. For one, since bubbles sample the Kolmogorov-like scales of vortex filaments, the resulting high spatial dishomogeneity emphasizes the action of the back-reaction force in the high wavenumber range. This warrants more caution in treating the dealiasing of the modified velocity field, as will be discussed in the next section. The second problem concerns the stability of the two-way coupled system, as discussed after.

\subsection{Dealiasing}\label{sec:dealiasing}

After that the particle feedback force $\bm{f}_p(\bm{x})$ is computed in real space on the $N^3$ grid, it is transformed into Fourier space, yielding $\hat{\bm{f}}_p(\bm{k})$ on the $N^3$ Fourier grid with $\bm{k} \in \left[-(N/2-1), ..., N/2\right]^3$. In a conventional pseudospectral algorithm with a 2/3-dealiasing rule, all Fourier space fields would be zeroed for all $|\bm{k}|>(2/3) N/2$ to avoid spurious aliasing contributions from the non-linear term.

\begin{figure}[t!]
    \centering
    \includegraphics[width=0.9\linewidth]{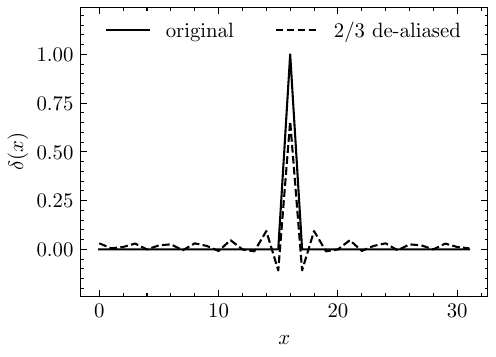}
    \caption{A 2/3-dealiasing rule applied to a $\delta$-function in real space reduces the magnitude of the $\delta$-peak and produces spurious oscillations in its vicinity.}
    \label{fig:dealiasing_example}
\end{figure}

However, in the case of light particles, due to their strong clustering, the particle feedback force becomes strongly localized and rough (large gradients), such that $\hat{\bm{f}}_p(\bm{k})$ has significant contributions in the high wavenumber range. Hence, truncating $\hat{\bm{f}}_p(\bm{k})$ using a 2/3-dealiasing rule would be a strong approximation, significantly decreasing the intensity of the two-way coupling, and it should thus be avoided. See Fig.~\ref{fig:dealiasing_example} for an example of the consequences of 2/3-dealiasing on a delta-function.

\begin{figure*}
    \centering
    \includegraphics[width=\linewidth]{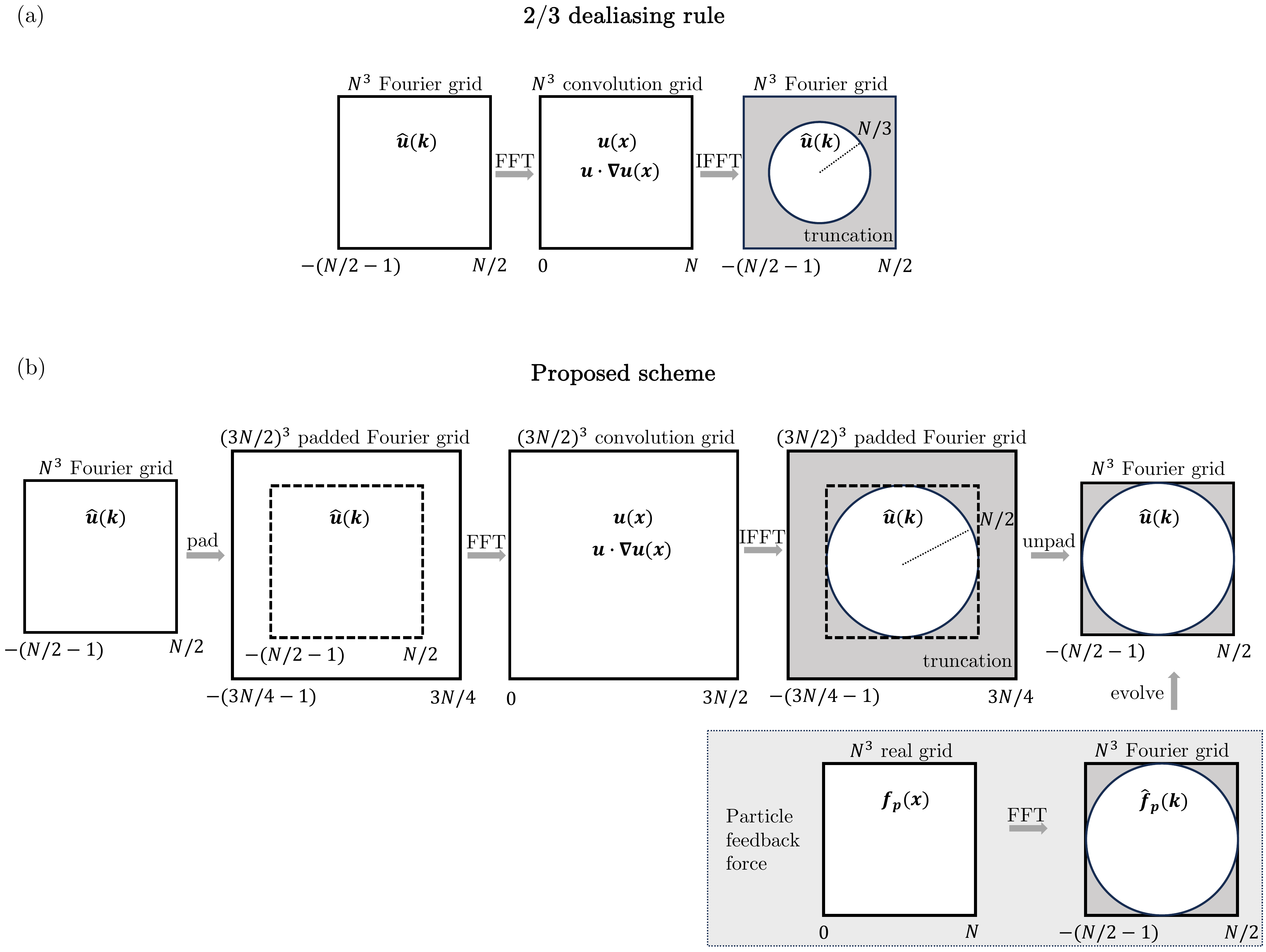}
    \caption{A schematic of the proposed dealiasing routine. The schematic depicts the analog 2D approach, but a 3D setting is implied. Panel (a): the conventional 2/3 dealiasing rule, where all quantities are computed on a $N^3$ grid and the Fourier space quantities are truncated at $|\bm{k}|>N/3$. Panel (b): our suggested routine. The flow field is evolved on a $N^3$ Fourier grid. For computation of the non-linear term, the flow field is inflated and padded with zeros to make a $(3N/2)^3$ grid, which is transformed to real space, where the convolution is carried out on the $(3N/2)^3$ convolution grid. The non-linear term is then taken back to Fourier space, where all contributions on wavenumbers $|\bm{k}|>N/2$ are removed for dealiasing. This yields a fully dealiased flow field on the $N^3$ Fourier grid. Other contributions such as the particle feedback force can be computed on a $N^3$ real space and $N^3$ Fourier space grid, needing only truncation of the wavenumbers $|\bm{k}|>N/2$, rather than $|\bm{k}|>N/3$ as in the conventional 2/3-dealiasing routine.}
    \label{fig:dealiasing_scheme}
\end{figure*}

Therefore, we propose a slightly different dealiasing approach where the non-linear term is computed on a larger grid than the one used for other real space quantities such as the feedback force, avoiding the need for truncation of the latter. The scheme is laid out in Fig.~\ref{fig:dealiasing_scheme}.

The zero-padded Fourier transform from Fourier space to the real space convolution grid and vice-versa can be computed efficiently as a pruned FFT and by skipping 1D transforms that are identically zero. This is implemented using the P3DFFT package \cite{Pekurovsky2012}. Note that the dealiasing approach suggested here raises the effective resolution of the simulation to $(3N/2)^3$ as compared to a $N^3$ resolution with a conventional 2/3-dealiasing rule.

An alternative way to mitigate the dealiasing problems would be to replace the delta-function by a broader smoother kernel over which the momentum is coupled back to the fluid. This can conveniently be done within the current scheme by replacing the delta function in Eq.~\eqref{eq:bubble_force} by a different compact function, although this will compromise the physical connection with vanishingly small point-particles.

\subsection{Stability}

The strong clustering of light particles not only poses a challenge due to the roughness that it introduces in the feedback force, but it can also compromise the stability of the integration of the two-way coupled system. This becomes apparent by considering Eq.~\eqref{eq:two-way_bubs}. Recall that the explicit time derivative $\partial \bm{u} / \partial t$ with respect to which the integration is performed is contained in the material derivative. Now, Eq.~\eqref{eq:two-way_bubs} shows that due to the two-way coupling, the time derivative is fed back into the evolution equation itself. Practically, since the material derivative as sensed by the particles is evaluated on one timestep $\Delta t$ prior to the current integration timestep, this effectively reduces Eq.~\eqref{eq:two-way_bubs} to
\begin{equation}
    \frac{\partial \bm{u}(t)}{\partial t} \approx \alpha\frac{\partial \bm{u}(t-\Delta t)}{\partial t} + \mathcal{F}[\bm{u}],
\end{equation}
with $\mathcal{F}[\bm{u}]$ capturing the other forces (pressure gradient, advective acceleration, viscosity and external force). This is a dynamical system of the type $b_{n} = \alpha b_{n-1}$, which becomes unstable if $\alpha > 1$, yielding exponentially growing solutions. This indicates that we should expect the two-way coupled system of bubbles to become unstable if the local volume fraction $\alpha$ exceeds unity. Physically, this can never occur if one enforces excluded volume interactions. This warrants the need for explicit treatment of the particle-particle interaction, indicating that in order to obtain reliable two-way coupled simulations of bubbly turbulence, one in fact needs to resort to four-way coupled simulations. This is treated in Sec. \ref{sec:part-part}.

The particle-particle interaction that will be considered here, however, is an approximate method in order to be able to computationally handle large number densities of bubbles. As a consequence, not all excluded volume interactions are strictly enforced at all times. To ensure stability, we therefore need to explicitly enforce that the local volume fraction does not exceed unity by carefully clipping the particle feedback force in those cases. To that extent, we multiply the particle feedback force by a local clipping factor $c(\bm{x},t)$ as
\begin{equation}
    c(\bm{x},t) = \begin{cases} 1 & \text{for } \alpha(\bm{x},t) \leq \alpha_0, \\ \frac{\alpha_0}{\alpha(\bm{x},t)} & \text{for } \alpha(\bm{x},t) > \alpha_0. \end{cases}
\end{equation}
By carefully clipping the particle feedback force in just the right places and times, this prefactor enforces that the effective local volume fraction never exceeds $\alpha_0$, ensuring the stability of the integration. While any $\alpha_0 < 1$ works, we set it to the close packing fraction of spheres $\alpha_0 = \pi / (3\sqrt{2}) \approx 0.74$.

For our benchmarking simulations of four-way coupled bubbly turbulence (bubbles with $D=0.8\eta$ at varying volume fraction $\bar{\alpha}$), we find that the total integrated magnitude of $\bm{f}_p$ (L1 norm) is clipped by less than 1\% as the volume fraction remains $\bar{\alpha} \lesssim 15\%$, which we deem acceptable.

\subsection{Performance}

The two-way coupling routine involves one loop over all Lagrangian particles as well as one FFT over the Eulerian grid. This routine can thus always be run within a runtime that is less than the combined runtime of the Eulerian part (involving two FFTs) and the integration of Lagrangian trajectories (involving also one loop over the Lagrangian particles, but with a heavier workload). Hence, the two-way coupling has a minimal impact on the overall performance.

\section{Particle-particle interaction} \label{sec:part-part}

\subsection{Efficient excluded volume interaction} \label{sec:excluded_volume}
To enforce the excluded volume hard-sphere interactions between particles, one needs to solve overlaps between particles at every time step. Solving all excluded volume interactions between all particles submerged in the fluid at large number densities is potentially a daunting problem from the computational point of view. However, instead of checking for all $N_p^2/2$ possible collisions between particles, we can use a boxing approach. This is a classical approach borrowed from Molecular Dynamics that only checks locally for possible collisions \cite{Allen1987}. Another computational difficulty is posed by particles that are involved in multiple collisions. Solving all overlaps recursively can in turn also become computationally intractable for sizable clusters of overlapping particles. We therefore propose an approximate algorithm for the excluded volume interaction, with efficiency in mind, that at every timestep solves only the strongest overlap for every particle. The second most severe overlap can then be solved in the next timestep, and so on. This allows us to have an excluded volume interaction algorithm that is efficient enough to allow for large number densities (with the number of particles on the same order as the number of Eulerian grid points), but also accurate enough to ensure a minimal total overlap volume of the particles. We coin this algorithm \textit{YOCO} ``You Only Collide Once''. The algorithm is laid out below.
\\\\
\begin{figure}[b!]
    \centering
    \includegraphics[width=0.8\linewidth]{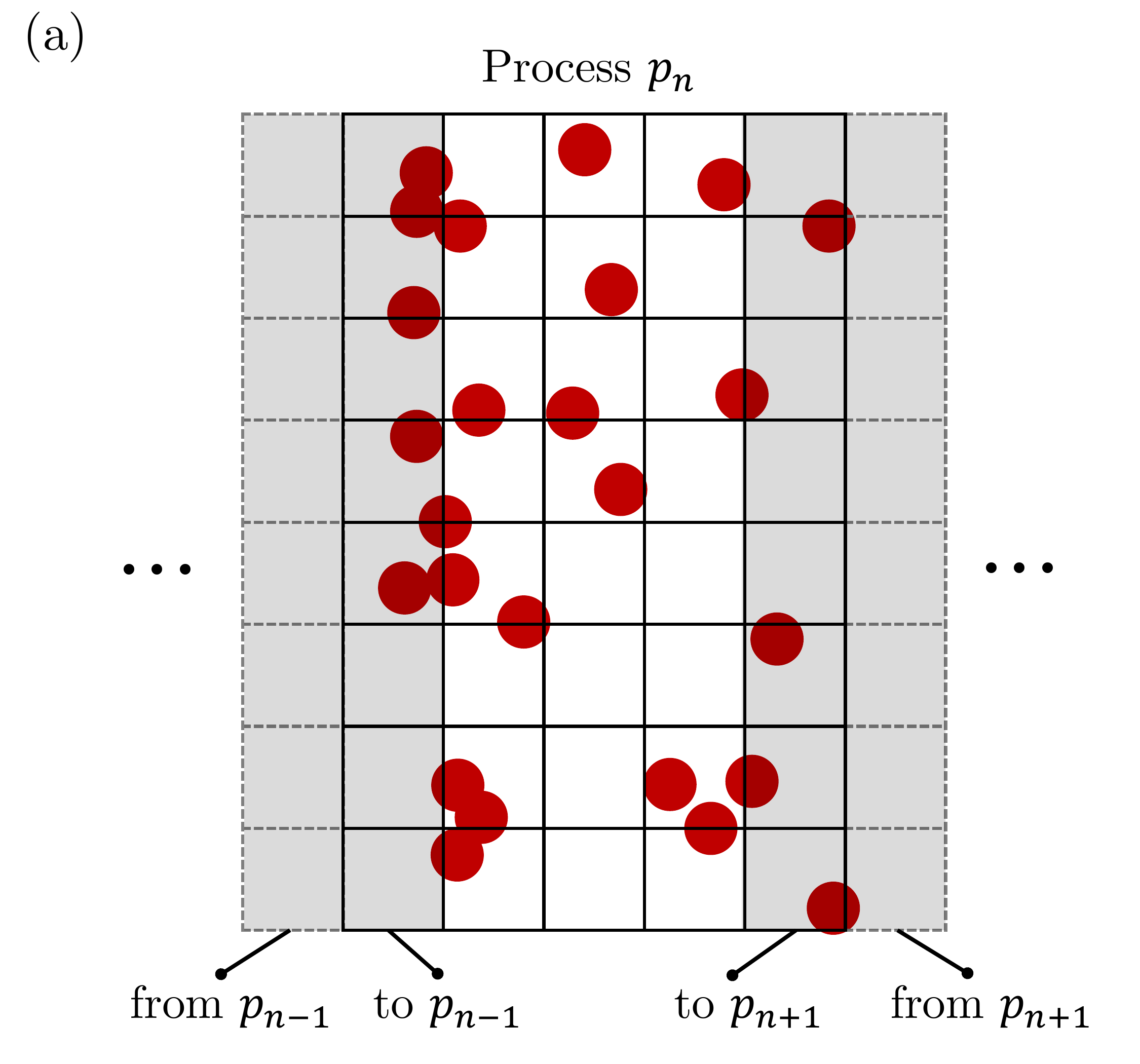}
    \caption{The boxing approach boxes all particles contained in process $p_n$ (solid boxes), while it sends (shaded solid boxes) and receives (shaded dashed boxes) the outer boxes to its nearest neighboring processes. Overlaps are then only checked between the particles contained within each box itself and its nearest neighboring boxes.}
    \label{fig:overlap_1}
\end{figure}
\begin{figure*}
    \centering
    \includegraphics[width=0.85\linewidth]{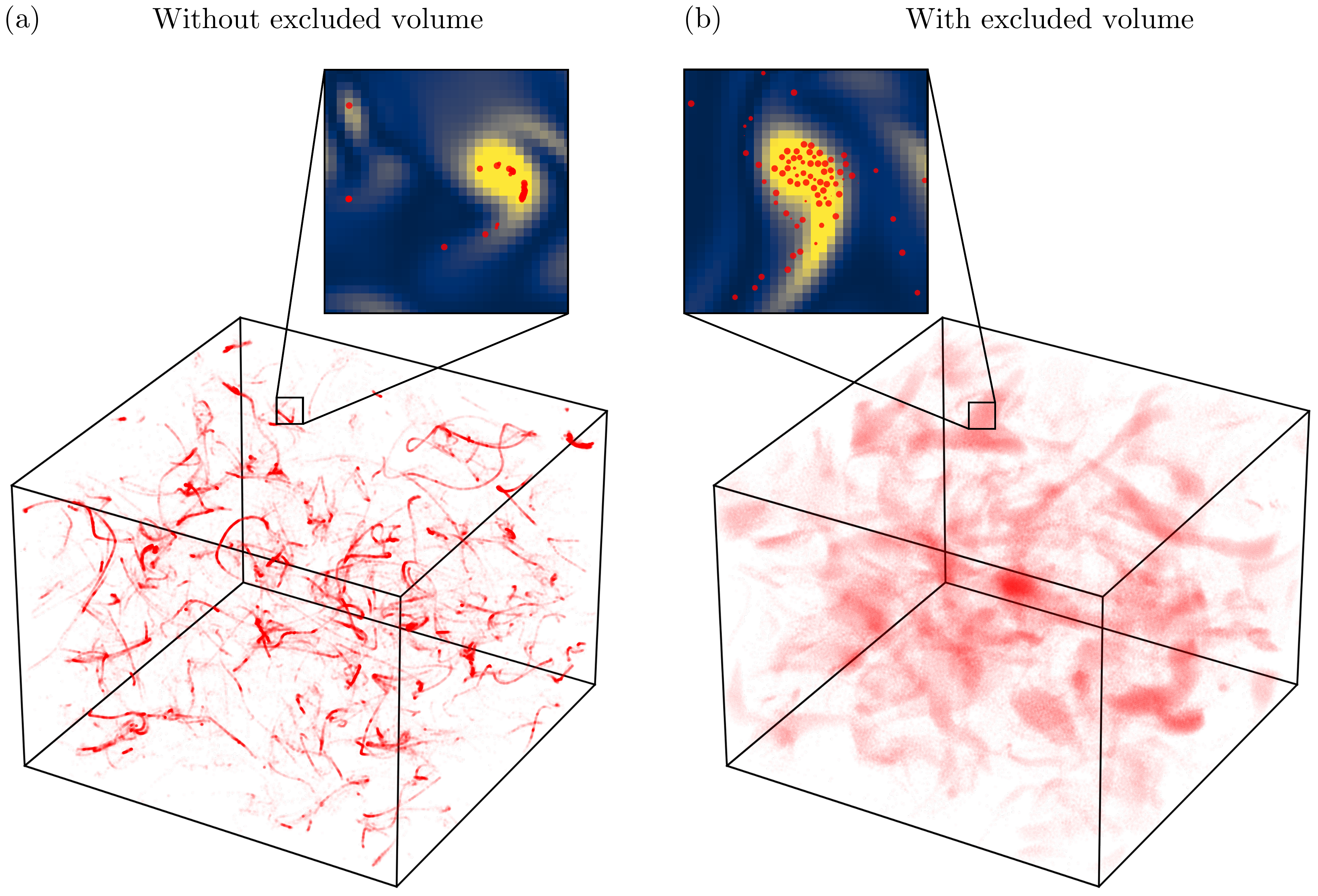}
    \caption{A snapshot of particle positions for a simulation without excluded volume interactions (a) and with excluded volume interactions (b) with bubbles of size $D=0.8\eta$ and average volume fraction $\bar{\alpha}=3.5\%$. The resolution is $N^3=128^3$ and number of particles $N_p=262\,144$ and Taylor scale Reynolds number $\textrm{Re}_\lambda\approx35$. The inset zooms show a small cross section of the full domain, where the background color indicates the local enstrophy $\Omega^2=|\bm{\nabla} \times \bm{u}|^2$.}
    \label{fig:particle_colls}
\end{figure*}

\begin{figure}[b!]
    \centering
    \includegraphics[width=0.8\linewidth]{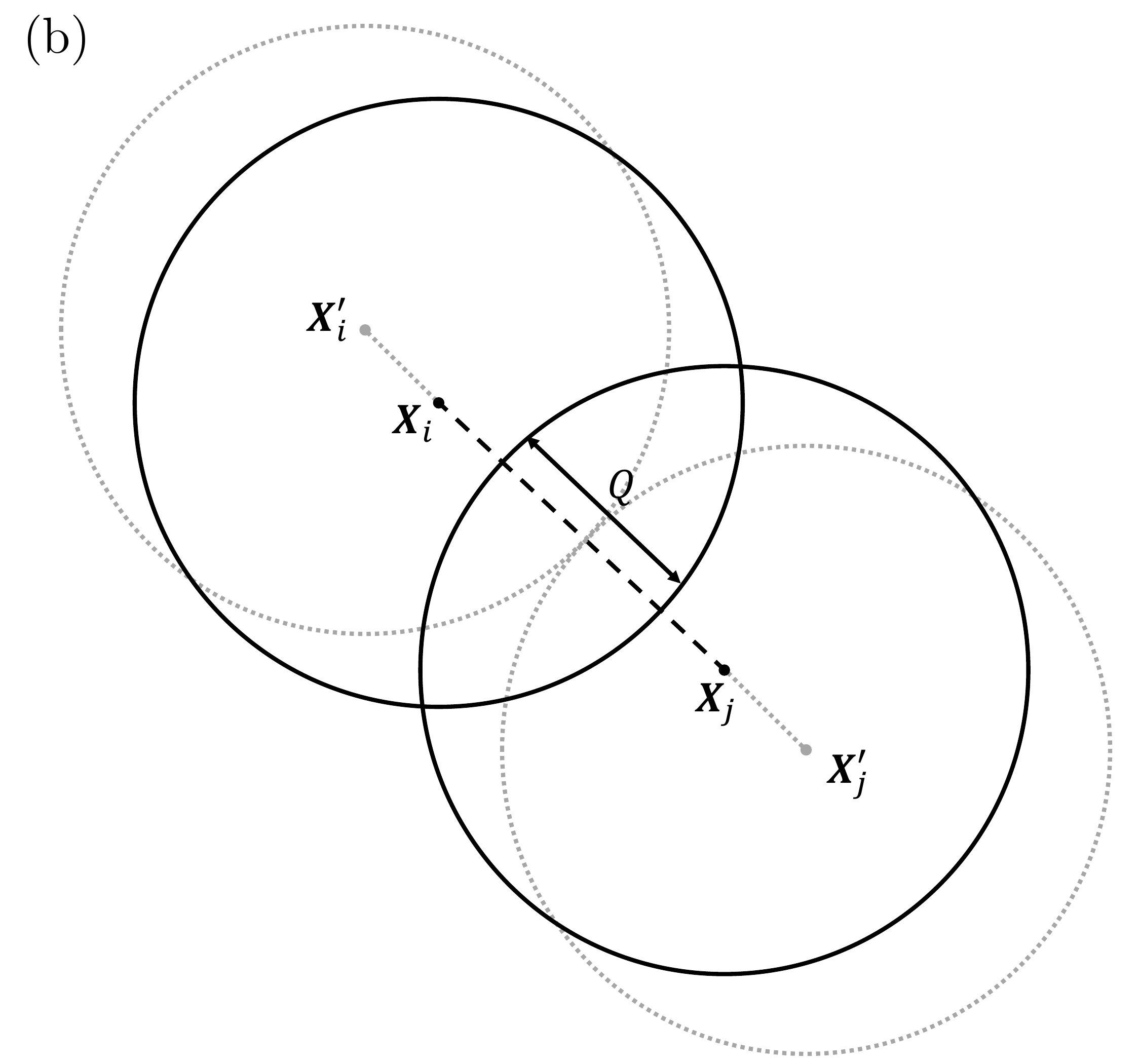}
    \caption{The excluded volume hard-sphere interactions are enforced by moving overlapping pairs of particles radially outwards over their mutual centerline from $\bm{X}_i$, $\bm{X}_j$ to $\bm{X}_i'$, $\bm{X}_j'$.}
    \label{fig:overlap_2}
\end{figure}

At every timestep:
\begin{enumerate}
    \item \textbf{Boxing}. The bubbles are collected in cubic boxes with sides that are minimally one diameter of a particle. This requires the creation of an array of boxes that tiles the full 3D space. By looping over all particles, this array is filled with references to the particles contained in that box. Each parallel process needs to communicate one extra layer of boxes in each direction to its neighboring processes to facilitate the overlap checking in the next step. See Fig.~\ref{fig:overlap_1}.
    \item \textbf{Overlap checking}. By looping over each box, we check for overlaps between particles within the box as well as between particles in the box and in the directly neighboring boxes. To avoid double counting, we check only for overlaps between particle $i$ and $j$ when the globally unique identifier of particle $i$ is larger than that of particle $j$. All overlaps are recorded in a local list, storing the references of the overlapping particles and the linear size of the overlap $Q=D-|\bm{X}_i - \bm{X_j}|$. We also check for \textit{forecasted collisions} that are predicted to happen within the next timestep based on the current velocities of the particles.
    \item \textbf{Sorting}. The list of overlaps is sorted from largest to smallest $Q$.
    \item \textbf{Purging}. We eliminate every second occurrence of a particle, leaving only the strongest overlap for each particle in the list. This can be done efficiently using a hash map as a look-up table to keep track of whether particles have already occurred in the list.
    \item \textbf{Solving collisions}. All remaining overlaps in the list are solved by moving each pair of overlapping particles outwards over their mutual centerline such that the new $Q=0$ for this pair as depicted in Fig~\ref{fig:overlap_2}. The \textit{forecasted collisions} are solved by performing an elastic collision on the particle velocities.
\end{enumerate}

To assess the effectiveness of this \textit{YOCO} algorithm in enforcing the most important excluded volume interactions, we perform a simulation of bubbly turbulence with and without excluded volume interaction. A resulting exemplary snapshot of particle positions is provided in Fig.~\ref{fig:particle_colls}. This shows that, indeed, due to excluded volume interactions, particles become notably more spread out around the vicinity of vortex filaments, rather than collapsing into the centers of the filaments.

\begin{figure*}
    \centering
    \includegraphics[width=0.85\linewidth]{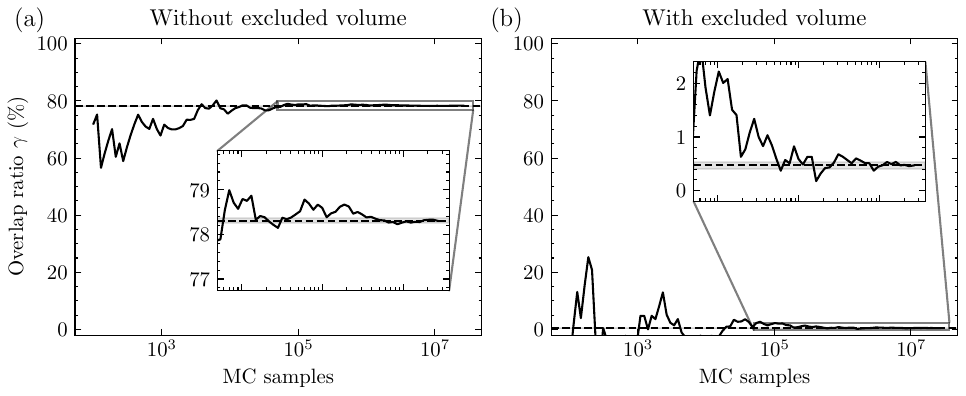}
    \caption{Monte-Carlo sampling of the overlap volume for a simulation with bubbles of size $D=0.8\eta$ and average volume fraction $\bar{\alpha}=3.5\%$ for the case without excluded volume interaction (a) and with excluded volume interaction (b) between bubbles. The figures show the convergence over the obtained overlap ratio as a function of the number of Monte-Carlo samples taken.}
    \label{fig:MC_volume}
\end{figure*}

\begin{figure}[b!]
    \centering
    \includegraphics[width=0.9\linewidth]{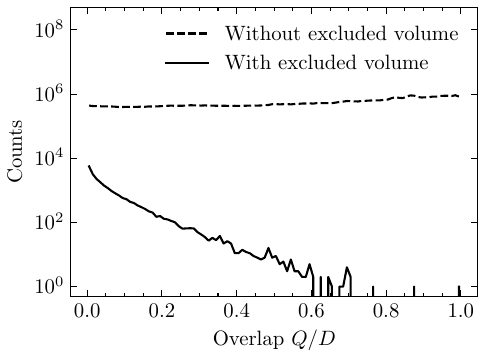}
    \caption{Histogram of the relative overlap $Q/D$ of the found overlapping pairs in a snapshot of a simulation without excluded volume interaction (dashed line) and with excluded volume interaction (solid line) for a simulation with bubbles of size $D=0.8\eta$ and average volume fraction $\bar{\alpha}=3.5\%$.}
    \label{fig:Overlap_hist}
\end{figure}

At the more quantitative level, we can compute the overlap volume by Monte-Carlo sampling probes in the whole space and checking whether they are inside or outside at least one particle. The fraction of Monte-Carlo samples that are inside at least one particle will then converge to $V_u/V$ with $V_u$ the total union volume of all particles and $V$ the total volume of the simulation domain. In the case of perfect excluded volume interactions, one should obtain $V_u \to \sum_i^{N_p} \mathcal{V}_p$, the summed volume of all particles. We can thus define the overlap ratio as
\begin{equation}
    \gamma = \frac{\sum_i^{N_p} \mathcal{V}_p - V_u}{\sum_i^{N_p} \mathcal{V}_p},
\end{equation}
which yields $\gamma\to 0$ in the case of perfect excluded volume interactions and $\gamma\to 1$ in the case of total overlap (vanishing union volume). The results of this Monte-Carlo sampling are presented in Fig.~\ref{fig:MC_volume}, which shows that the overlap ratio for the simulation without excluded volume interaction yields $\gamma=(78.3 \pm 0.1)\%$ while in the simulation with excluded volume interactions an overlap ratio $\gamma=(0.5 \pm 0.1)\%$ remains. This shows that the \textit{YOCO} algorithm is very effective in reducing the overlap between particles, with an almost negligible remaining overlap ratio. The histogram of the found overlaps for the cases with and without excluded volume interactions is provided in Fig.~\ref{fig:Overlap_hist}. This shows that for the case with excluded volume interactions, most remaining overlaps are small, while for the case without excluded volume interactions, the most probable overlap is close to full overlap $Q/D\approx1$. This further assures us of the effectiveness of the proposed \textit{YOCO} algorithm. We find that when pushing the volume fraction $\bar{\alpha}$ even further into the very dense regime, the overlap ratio remains $\gamma \lesssim 3\%$ while $\bar{\alpha} \lesssim 15\%$.

\subsection{Evaluation of the material derivative}

A final point of attention concerns the evaluation of the material derivative at the position of the particle as needed for the particle equation of motion Eq.~\eqref{eq:particle}. Recall that the material derivative at the position of the particle is given as
\begin{equation}\label{eq:mat_der_pure}
    \frac{\mathrm{D}\bm{u}(\bm{X},t)}{\mathrm{D}t} = \frac{\partial\bm{u}(\bm{X},t)}{\partial t} + \bm{u}(\bm{X},t)\cdot\bm{\nabla}\bm{u}(\bm{X},t).
\end{equation}
A commonly used trick in Lagrangian tracking is to evaluate the material derivative at the particle position from the full derivative of the fluid velocity at the position of the particle $\mathrm{d}\bm{u}(\bm{X},t)/\mathrm{d}t$. This quantity is numerically easily accessible, as it involves only the result of the interpolation of the velocity at the previous timestep. Then, considering that
\begin{equation}
    \frac{\mathrm{d}\bm{u}(\bm{X},t)}{\mathrm{d}t} = \frac{\partial \bm{u}(\bm{X},t)}{\partial t} + \frac{\mathrm{d}\bm{X}(t)}{\mathrm{d}t}\cdot\bm{\nabla}\bm{u}(\bm{X},t),
\end{equation}
the material derivative at the position of the particle can be obtained as
\begin{equation}\label{eq:mat_der_trick}
    \frac{\mathrm{D}\bm{u}(\bm{X},t)}{\mathrm{D}t} = \frac{\mathrm{d}\bm{u}(\bm{X},t)}{\mathrm{d}t} + \left[\bm{u}(\bm{X},t) - \bm{V}(t)\right]\cdot\bm{\nabla}\bm{u}(\bm{X},t).
\end{equation}
This conveniently avoids the evaluation of the partial time derivative of the fluid velocity at the current position of the particle, which would require knowledge of the previous state of the full flow field. However, note that here we have tacitly assumed that $\mathrm{d}\bm{X}(t)/\mathrm{d}t = \bm{V}(t)$. While this is usually true, this is not the case with our treatment of particle-particle collisions, which can alter the particle position due to collisions in a way that does not follow the integration of the velocity alone.

Therefore, when applying particle-particle interactions, we do not evaluate the material derivative at the particle position according to Eq.~\eqref{eq:mat_der_trick}, but we instead evaluate Eq.~\eqref{eq:mat_der_pure} directly. This involves keeping one previous flow snapshot in memory and interpolating both the current velocity as well as the previous velocity at the current position of the particle in order to evaluate the partial time derivative. This can be done with a minimal impact on the overall performance of the code.

\subsection{Performance}
\begin{figure}[t!]
    \centering
    \includegraphics[width=0.95\linewidth]{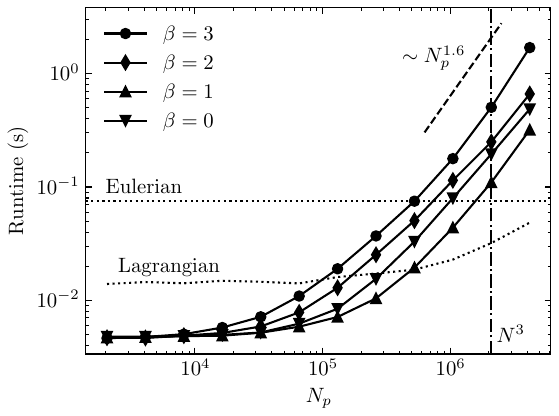}
    \caption{Performance of the four-way coupling method as a function of the number of particles $N_p$. Solid lines with symbols denote the runtime of the proposed \textit{YOCO} algorithm for the particle-particle interaction for particles with different buoyancy ratios. The horizontal dotted line/curve coined Eulerian and Lagrangian denote the runtime of routines for the Eulerian part and of the integration of the Lagrangian trajectories, respectively. The vertical dashed-dotted line denotes the number of Eulerian grid points $N^3=128^3$ for reference. The particle size $D=0.65\eta$, such that the volume fraction varies from $\bar{\alpha}=0.01 \%$ to $\bar{\alpha}=29 \%$.}
    \label{fig:profiling}
\end{figure}

Treating pairwise particle collision is a problem that can in principle be of quadratic complexity in the number of particles. While the algorithm proposed here effectively reduces the complexity, for example through the boxing approach that employs the locality of collisions, the complexity of the algorithm is still expected to exceed linearity. For an increasingly large number of particles, the collision routine can thus become the computationally heaviest part of the full four-way coupling code. We have therefore carefully assessed the performance of the four-way coupling method as laid out in Fig.~\ref{fig:profiling}. Since the performance of the particle-particle routine depends on the number of realized collisions, the slowest runtime is obtained for bubbles ($\beta=3$), as expected. However, as can be appreciated from the figure, thanks to the efficient implementation of the \textit{YOCO} algorithm, the collision algorithm only starts to bottleneck the overall performance as the number of particles approaches the number of Eulerian grid points ($N^3$). Indeed, here we obtain a scaling of the runtime with the number of particles with a power between 1 and 2.

\section{Example of four-way coupling}\label{sec:example}
\begin{figure*}
    \centering
    \includegraphics[width=0.85\linewidth]{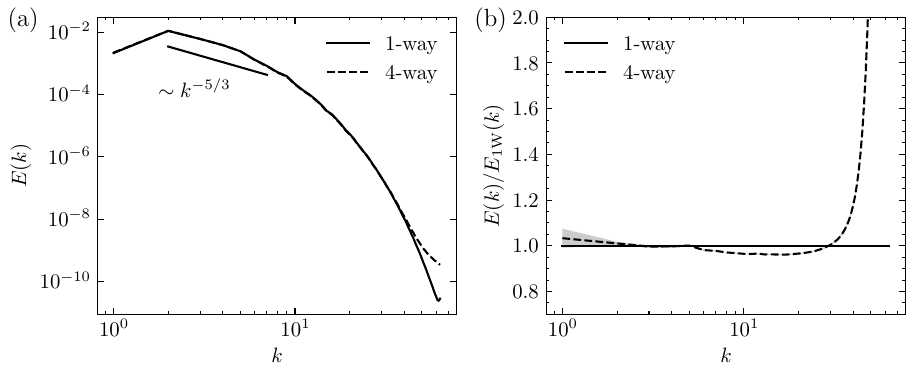}
    \caption{Panel (a): Energy spectra of the one-phase turbulent flow and the four-way coupled bubbly turbulent flow with average volume fraction of bubbles $\bar{\alpha}=3.5\%$ with size $D=0.8\eta$. Panel (b): Four-way coupled energy spectrum compensated by the one-way energy spectrum. This shows a slight attenuation of the turbulence at intermediate to dissipative scales and a relatively strong enhancement of the smallest scales. The shaded regions denote statistical errorbars. The resolution is $N^3=128^3$ and number of particles $N_p=262\,144$ and Taylor scale Reynolds number $\textrm{Re}_\lambda\approx35$. Compare with Fig.~\ref{fig:spectra_2w_bench} that is run at enforced uniform particle density.}
    \label{fig:spectra_4w_bench}
\end{figure*}

Finally, we present an example simulation of the full four-way coupled bubbly turbulence in Fig.~\ref{fig:spectra_4w_bench}. At contrast with the artificial test case where bubbles were kept fixed to enforce uniform density as presented in Fig.~\ref{fig:spectra_2w_bench}, here we see that the amplification of the large and inertial scales vanishes. Instead, when the bubbles are free to move and preferentially concentrate, we observe a slight attenuation of the intermediate to dissipative scales and a strong amplification of the smallest scales. We argue that the attenuation is an effect of increased effective viscosity, while the amplification of the smallest scales can be understood as a redistribution of energy towards the smallest scales where the particles are active. These findings are qualitatively consistent with earlier experimental and numerical findings \cite{Lance1991,Riboux2010,Almeras2017,Mazzitelli2003}. In future work, we plan to assess the effect of four-way coupling further using the method presented here, studying e.g. the effect on the preferential concentration, pair dispersion, and energy spectra under the influence of different concentrations of various particles and for various Reynolds numbers.

\section{Conclusions and outlook}\label{sec:conclusion}
In this work, we have presented a four-way coupling approach for simulating small particles in a turbulent flow, tailored towards supporting high number densities of particles and/or cases of intense particle clustering. While more accurate methods exist, which extend the support of particle sizes up to the integral scales of turbulence, these more complex methods compromise on computational efficiency, limiting the number of particles that can be resolved. Our point-based method, on the other hand, has a computational cost that is of a similar order as the one for the evolution of the Eulerian flow, at least when the number of particles remains of the same order of the number of Eulerian grid points, which in practice amounts to millions of particles. This method is thus well suited for studying the collective effect of large volume fractions of small submerged particles in turbulent flows as encountered in plenty of industrial and natural applications \cite{Brandt2022}.

Moreover, our method can be straightforwardly extended to more complicated systems, for example when there are other forces acting on the particle, such as an external driving force. By extending the equation of motion of the particle, the feedback mechanism presented here automatically couples its dynamics to the dynamics of the background turbulence through the mutual momentum exchange. The method presented here can also be combined with other grid-based flow solvers, which would allow the treatment of e.g. wall-bounded turbulence. Another extension that one can make is to include rotational dynamics of the particle as well as its feedback onto the flow, as is laid out in Ref.~\cite{Andersson2012}. The versatility of this method thus opens up many opportunities to study a large variety of complex particle-laden fluid systems in which resolving a large number of particles is important to capture their collective physical effect.

\section*{Acknowledgements}
We thank Dmitry Pekurovsky for the development of the P3DFFT library. This publication is part of the project “Shaping turbulence with smart particles” with project number OCENW.GROOT.2019.031 of the research programme Open Competitie ENW XL which is (partly) financed by the Dutch Research Council (NWO).

\appendix
\section{Scheme for Eulerian pseudospectral solver}\label{sec:eulerian_solver}
For the Eulerian part, a classical pseudospectral approach is employed. To solve the Navier-Stokes equations
\begin{equation}\label{eq:NS_app}
    \frac{\partial \bm{u}}{\partial t} + \bm{u} \cdot \bm{\nabla} \bm{u} = -\bm{\nabla} p + \nu \Delta \bm{u} + \bm{f},
\end{equation}
under the condition of incompressibilty $\bm{\nabla}\cdot\bm{u}=0$, we write the velocity field $\bm{u}$ in terms of its vector potential $\bm{b}$ as
\begin{equation}
    \bm{u} = \bm{\nabla} \times \bm{b},
\end{equation}
which inherently ensures incompressibility since \mbox{$\bm{\nabla}\cdot\bm{\nabla} \times \bm{b}=0$}.

Taking the curl of Eq.~\eqref{eq:NS_app} and using the definition of vorticity $\bm{\omega}=\bm{\nabla}\times\bm{u}$, then yields the Navier-Stokes equations in vorticity-velocity formulation
\begin{equation}
    \frac{\partial \bm{\omega}}{\partial t} = \bm{\nabla}\times(\bm{u}\times\bm{\omega}) + \nu \Delta \bm{\omega} + \bm{\nabla}\times \bm{f}.
\end{equation}
This can be rewritten as a dynamical equation for the vector potential by noting that $\bm{\omega}=-\Delta\bm{b}$ as
\begin{equation}\label{eq:interm}
    -\frac{\partial \Delta\bm{b}}{\partial t} = \bm{\nabla}\times(\bm{u}\times\bm{\omega}) - \nu \Delta^2 \bm{b} + \bm{\nabla}\times \bm{f}.
\end{equation}
This equation is solved in spectral $\bm{k}$-space, where we define the Fourier transform $\hat{\bm{b}}(\bm{k})=\mathcal{F}[\bm{b}(\bm{x})]_{\bm{k}}$. Then we can write Eq.~\eqref{eq:interm} in the final pseudospectral form
\begin{equation}\label{eq:PS}
\boxed{
    |\bm{k}|^2 \frac{\partial \hat{\bm{b}}}{\partial t} = i\bm{k}\times\mathcal{F}\left[\bm{u}\times\bm{\omega}\right]_{\bm{k}} - \nu |\bm{k}|^4\hat{\bm{b}} + i\bm{k}\times \hat{\bm{f}}.
}
\end{equation}
In practice, this equation is evaluated on a finite $\bm{k}$-space of size $N^3$. This should be large enough to resolve the smallest active scale, being the dissipative Kolmogorov scale $\eta$, such that the largest wavenumber $k_{\textrm{max}}\approx2\pi/\eta$.
The non-linear term $\mathcal{F}\left[\bm{u}\times\bm{\omega}\right]_{\bm{k}}$ is evaluated in real space by first computing the velocity and vorticity in real space as 
\begin{equation}
\bm{u}=\mathcal{F}^{-1}\left[i\bm{k}\times\hat{\bm{b}}\right]_{\bm{x}}, \quad  \bm{\omega}=\mathcal{F}^{-1}\left[|\bm{k}|^2 \hat{\bm{b}}\right]_{\bm{x}}.
\end{equation}
and then carrying out the cross product between them using pointwise multiplication and transforming this product back to the Fourier space. This is the essence of the pseudospectral method, as evaluating the non-linear term in spectral space would involve a convolution of quadratic complexity, while the Fourier transform can be performed with lin-log complexity, owing to the Fast Fourier Transform (FFT) algorithm \cite{FFTW05}. In 3D on an $N^3$ Eulerian grid, the complexity of the pseudospectral method thus becomes $\mathcal{O}(N^3\log N)$. For an efficient parallel implementation of the FFT in 3D, we use the P3DFFT package \cite{Pekurovsky2012}.

The pseudospectral method produces higher harmonics in the non-linear term that need to be dealiased. This dealiasing needs to be done after the non-linear term is transformed back into the Fourier space. Our dealiasing approach is laid out in the main text in Sec. \ref{sec:dealiasing}.

To solve Eq.~\eqref{eq:PS}, we need to integrate it in time. Let us rewrite it as
\begin{equation}
    \frac{\partial \hat{\bm{b}}(\bm{k},t)}{\partial t} = c \hat{\bm{b}}(\bm{k},t) + K(\hat{\bm{b}}(\bm{k},t),t),
\end{equation}
where $c\equiv -\nu|\bm{k}|^2$ and $K(\hat{\bm{b}}(\bm{k},t),t) \equiv |\bm{k}|^{-2}(i\bm{k}\times\mathcal{F}\left[\bm{u}\times\bm{\omega}\right]_{\bm{k}} + i\bm{k}\times \hat{\bm{f}})$. Then, we can solve for the viscosity exactly by treating it as an integrating factor
\begin{equation}
    \frac{\partial}{\partial t}\left[\hat{\bm{b}}(\bm{k},t) \exp(-ct) \right] = K(\hat{\bm{b}}(\bm{k},t),t) \exp(-ct).
\end{equation}
Integrating for a small timestep $dt$ from $t_n$ to \mbox{$t_{n+1}=t_n+dt$} then gives
\begin{eqnarray}
    \hat{\bm{b}}(\bm{k},&&t_{n+1}) = \hat{\bm{b}}(\bm{k},t_n)\exp(cdt) \\ &&+ \exp(cdt) \int_0^{dt} \exp(-c\tau) K(\hat{\bm{b}}(\bm{k},t_n+\tau),t_n+\tau)\mathrm{d}\tau \nonumber.
\end{eqnarray}
To evaluate the temporal integral numerically, we employ a second-order accurate Adam-Bashfort scheme, yielding
\begin{eqnarray}
    \hat{\bm{b}}(\bm{k},t_{n+1}) \approx \hat{\bm{b}}(\bm{k},t_n)&&\exp(cdt)+ \left[ \frac{3dt}{2}K_n \exp(cdt) \right] \nonumber\\ && - \left[ \frac{dt}{2}K_{n-1} \exp(2cdt) \right],
\end{eqnarray}
where we abbreviated $K_{n} \equiv K(\hat{\bm{b}}(\bm{k},t_n),t_n)$ and $K_{n-1} \equiv K(\hat{\bm{b}}(\bm{k},t_{n-1}),t_{n-1})$. This completes the implementation of the Eulerian part of the numerical method presented here.

\section{Validation of the absence of self-induced motion}\label{sec:self_ind}

Here, we follow the approach that is put forward in Ref.~\cite{Mazzitelli2003} to verify that the self-induced motion created by the two-way coupling mechanism is not influencing the dynamics of the particles significantly. Formally, the Maxey-Riley equations Eq.~\eqref{eq:particle} assume that the flow field is perturbed by all other particles except for the particle under consideration itself. Since it is practically impossible to separate the contributions from the different disturbance fields created by each particle, we have to assume that the perturbation created by one particle to the flow field is only negligibly influencing its own dynamics.

To verify this, we compare the dynamics of particles that are back-coupled to the fluid to particles that are not back-coupled to the fluid, which are evolved in the same simulation. The coupled particles are thus influenced by the perturbations from all other coupled particles as well as from themselves, while the uncoupled particles are influenced by the perturbations from all coupled particles but not from themselves. All particles have the particle-particle excluded volume interactions as considered in the main text.

\begin{figure}[b!]
    \centering
    \includegraphics[width=0.95\linewidth]{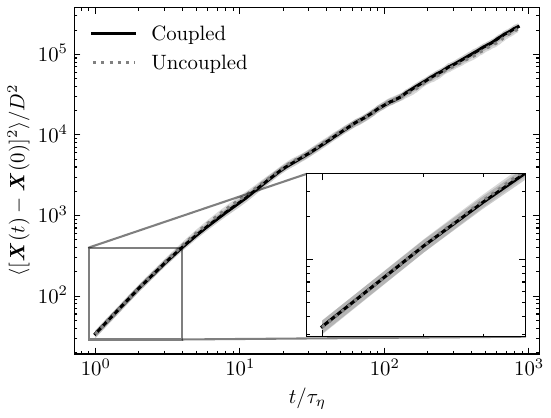}
    \caption{Comparison of the mean squared displacement of particles that are back-coupled to the fluid (solid line) and particles that are not back-coupled to the fluid (dotted line) evolved in the same simulation. The shaded region indicates the statistical uncertainty. All particles have excluded volume interactions. The particles are bubbles of size $D=0.8\eta$ and the average volume fraction $\bar{\alpha}=3.5\%$. The resolution is $N^3=128^3$ and number of particles $N_p=262\,144$ (half coupled and half uncoupled) and Taylor scale Reynolds number $\textrm{Re}_\lambda\approx35$.}
    \label{fig:self_ind_validation}
\end{figure}

In Fig.~\ref{fig:self_ind_validation}, we compare the diffusion of both classes of particles through their mean-squared displacement $\langle [\bm{X}(t)-\bm{X}(0)]^2 \rangle$. This shows that both classes of particles diffuse indistinguishably from one another, indicating that there is no measurable effect of the self-induced motion.

%\bibliography{main}

\begin{thebibliography}{39}%
\makeatletter
\providecommand \@ifxundefined [1]{%
 \@ifx{#1\undefined}
}%
\providecommand \@ifnum [1]{%
 \ifnum #1\expandafter \@firstoftwo
 \else \expandafter \@secondoftwo
 \fi
}%
\providecommand \@ifx [1]{%
 \ifx #1\expandafter \@firstoftwo
 \else \expandafter \@secondoftwo
 \fi
}%
\providecommand \natexlab [1]{#1}%
\providecommand \enquote  [1]{``#1''}%
\providecommand \bibnamefont  [1]{#1}%
\providecommand \bibfnamefont [1]{#1}%
\providecommand \citenamefont [1]{#1}%
\providecommand \href@noop [0]{\@secondoftwo}%
\providecommand \href [0]{\begingroup \@sanitize@url \@href}%
\providecommand \@href[1]{\@@startlink{#1}\@@href}%
\providecommand \@@href[1]{\endgroup#1\@@endlink}%
\providecommand \@sanitize@url [0]{\catcode `\\12\catcode `\$12\catcode `\&12\catcode `\#12\catcode `\^12\catcode `\_12\catcode `\%12\relax}%
\providecommand \@@startlink[1]{}%
\providecommand \@@endlink[0]{}%
\providecommand \url  [0]{\begingroup\@sanitize@url \@url }%
\providecommand \@url [1]{\endgroup\@href {#1}{\urlprefix }}%
\providecommand \urlprefix  [0]{URL }%
\providecommand \Eprint [0]{\href }%
\providecommand \doibase [0]{https://doi.org/}%
\providecommand \selectlanguage [0]{\@gobble}%
\providecommand \bibinfo  [0]{\@secondoftwo}%
\providecommand \bibfield  [0]{\@secondoftwo}%
\providecommand \translation [1]{[#1]}%
\providecommand \BibitemOpen [0]{}%
\providecommand \bibitemStop [0]{}%
\providecommand \bibitemNoStop [0]{.\EOS\space}%
\providecommand \EOS [0]{\spacefactor3000\relax}%
\providecommand \BibitemShut  [1]{\csname bibitem#1\endcsname}%
\let\auto@bib@innerbib\@empty
%</preamble>
\bibitem [{\citenamefont {Balachandar}\ and\ \citenamefont {Eaton}(2010)}]{Balachandar2010}%
  \BibitemOpen
  \bibfield  {author} {\bibinfo {author} {\bibfnamefont {S.}~\bibnamefont {Balachandar}}\ and\ \bibinfo {author} {\bibfnamefont {J.~K.}\ \bibnamefont {Eaton}},\ }\bibfield  {title} {\bibinfo {title} {Turbulent {Dispersed} {Multiphase} {Flow}},\ }\href {https://doi.org/10.1146/ANNUREV.FLUID.010908.165243} {\bibfield  {journal} {\bibinfo  {journal} {Annual Review of Fluid Mechanics}\ }\textbf {\bibinfo {volume} {42}},\ \bibinfo {pages} {111} (\bibinfo {year} {2010})}\BibitemShut {NoStop}%
\bibitem [{\citenamefont {Toschi}\ and\ \citenamefont {Bodenschatz}(2009)}]{Toschi2009}%
  \BibitemOpen
  \bibfield  {author} {\bibinfo {author} {\bibfnamefont {F.}~\bibnamefont {Toschi}}\ and\ \bibinfo {author} {\bibfnamefont {E.}~\bibnamefont {Bodenschatz}},\ }\bibfield  {title} {\bibinfo {title} {Lagrangian properties of particles in turbulence},\ }\href {https://doi.org/10.1146/ANNUREV.FLUID.010908.165210} {\bibfield  {journal} {\bibinfo  {journal} {Annual Review of Fluid Mechanics}\ }\textbf {\bibinfo {volume} {41}},\ \bibinfo {pages} {375} (\bibinfo {year} {2009})}\BibitemShut {NoStop}%
\bibitem [{\citenamefont {Mathai}\ \emph {et~al.}(2020)\citenamefont {Mathai}, \citenamefont {Lohse},\ and\ \citenamefont {Sun}}]{Mathai2020}%
  \BibitemOpen
  \bibfield  {author} {\bibinfo {author} {\bibfnamefont {V.}~\bibnamefont {Mathai}}, \bibinfo {author} {\bibfnamefont {D.}~\bibnamefont {Lohse}},\ and\ \bibinfo {author} {\bibfnamefont {C.}~\bibnamefont {Sun}},\ }\bibfield  {title} {\bibinfo {title} {Bubbly and {Buoyant} {Particle}–{Laden} {Turbulent} {Flows}},\ }\href {https://doi.org/10.1146/ANNUREV-CONMATPHYS-031119-050637} {\bibfield  {journal} {\bibinfo  {journal} {Annual Review of Condensed Matter Physics}\ }\textbf {\bibinfo {volume} {11}},\ \bibinfo {pages} {529} (\bibinfo {year} {2020})}\BibitemShut {NoStop}%
\bibitem [{\citenamefont {Brandt}\ and\ \citenamefont {Coletti}(2022)}]{Brandt2022}%
  \BibitemOpen
  \bibfield  {author} {\bibinfo {author} {\bibfnamefont {L.}~\bibnamefont {Brandt}}\ and\ \bibinfo {author} {\bibfnamefont {F.}~\bibnamefont {Coletti}},\ }\bibfield  {title} {\bibinfo {title} {Particle-laden turbulence: Progress and perspectives},\ }\href {https://doi.org/10.1146/ANNUREV-FLUID-030121-021103} {\bibfield  {journal} {\bibinfo  {journal} {Annual Review of Fluid Mechanics}\ }\textbf {\bibinfo {volume} {54}},\ \bibinfo {pages} {159} (\bibinfo {year} {2022})}\BibitemShut {NoStop}%
\bibitem [{\citenamefont {Benzi}\ and\ \citenamefont {Toschi}(2023)}]{Benzi2023}%
  \BibitemOpen
  \bibfield  {author} {\bibinfo {author} {\bibfnamefont {R.}~\bibnamefont {Benzi}}\ and\ \bibinfo {author} {\bibfnamefont {F.}~\bibnamefont {Toschi}},\ }\bibfield  {title} {\bibinfo {title} {Lectures on turbulence},\ }\href {https://doi.org/https://doi.org/10.1016/j.physrep.2023.05.001} {\bibfield  {journal} {\bibinfo  {journal} {Physics Reports}\ }\textbf {\bibinfo {volume} {1021}},\ \bibinfo {pages} {1} (\bibinfo {year} {2023})},\ \bibinfo {note} {lectures on turbulence}\BibitemShut {NoStop}%
\bibitem [{\citenamefont {Maxey}(1987)}]{Maxey1987}%
  \BibitemOpen
  \bibfield  {author} {\bibinfo {author} {\bibfnamefont {M.~R.}\ \bibnamefont {Maxey}},\ }\bibfield  {title} {\bibinfo {title} {The motion of small spherical particles in a cellular flow field},\ }\href {https://doi.org/10.1063/1.866206} {\bibfield  {journal} {\bibinfo  {journal} {The Physics of Fluids}\ }\textbf {\bibinfo {volume} {30}},\ \bibinfo {pages} {1915} (\bibinfo {year} {1987})}\BibitemShut {NoStop}%
\bibitem [{\citenamefont {Crisanti}\ \emph {et~al.}(1992)\citenamefont {Crisanti}, \citenamefont {Falcioni}, \citenamefont {Provenzale}, \citenamefont {Tanga},\ and\ \citenamefont {Vulpiani}}]{Crisanti1992}%
  \BibitemOpen
  \bibfield  {author} {\bibinfo {author} {\bibfnamefont {A.}~\bibnamefont {Crisanti}}, \bibinfo {author} {\bibfnamefont {M.}~\bibnamefont {Falcioni}}, \bibinfo {author} {\bibfnamefont {A.}~\bibnamefont {Provenzale}}, \bibinfo {author} {\bibfnamefont {P.}~\bibnamefont {Tanga}},\ and\ \bibinfo {author} {\bibfnamefont {A.}~\bibnamefont {Vulpiani}},\ }\bibfield  {title} {\bibinfo {title} {Dynamics of passively advected impurities in simple two‐dimensional flow models},\ }\href {https://doi.org/10.1063/1.858402} {\bibfield  {journal} {\bibinfo  {journal} {Physics of Fluids A: Fluid Dynamics}\ }\textbf {\bibinfo {volume} {4}},\ \bibinfo {pages} {1805} (\bibinfo {year} {1992})}\BibitemShut {NoStop}%
\bibitem [{\citenamefont {Balkovsky}\ \emph {et~al.}(2001)\citenamefont {Balkovsky}, \citenamefont {Falkovich},\ and\ \citenamefont {Fouxon}}]{Balkovsky2001}%
  \BibitemOpen
  \bibfield  {author} {\bibinfo {author} {\bibfnamefont {E.}~\bibnamefont {Balkovsky}}, \bibinfo {author} {\bibfnamefont {G.}~\bibnamefont {Falkovich}},\ and\ \bibinfo {author} {\bibfnamefont {A.}~\bibnamefont {Fouxon}},\ }\bibfield  {title} {\bibinfo {title} {Intermittent distribution of inertial particles in turbulent flows},\ }\href {https://doi.org/10.1103/PhysRevLett.86.2790} {\bibfield  {journal} {\bibinfo  {journal} {Physics Review Letters}\ }\textbf {\bibinfo {volume} {86}},\ \bibinfo {pages} {2790} (\bibinfo {year} {2001})}\BibitemShut {NoStop}%
\bibitem [{\citenamefont {Calzavarini}\ \emph {et~al.}(2008)\citenamefont {Calzavarini}, \citenamefont {Kerscher}, \citenamefont {Lohse},\ and\ \citenamefont {Toschi}}]{Calzavarini2008}%
  \BibitemOpen
  \bibfield  {author} {\bibinfo {author} {\bibfnamefont {E.}~\bibnamefont {Calzavarini}}, \bibinfo {author} {\bibfnamefont {M.}~\bibnamefont {Kerscher}}, \bibinfo {author} {\bibfnamefont {D.}~\bibnamefont {Lohse}},\ and\ \bibinfo {author} {\bibfnamefont {F.}~\bibnamefont {Toschi}},\ }\bibfield  {title} {\bibinfo {title} {Dimensionality and morphology of particle and bubble clusters in turbulent flow},\ }\href {https://doi.org/10.1017/S0022112008001936} {\bibfield  {journal} {\bibinfo  {journal} {Journal of Fluid Mechanics}\ }\textbf {\bibinfo {volume} {607}},\ \bibinfo {pages} {13} (\bibinfo {year} {2008})}\BibitemShut {NoStop}%
\bibitem [{\citenamefont {Peskin}(1972)}]{Peskin1972}%
  \BibitemOpen
  \bibfield  {author} {\bibinfo {author} {\bibfnamefont {C.~S.}\ \bibnamefont {Peskin}},\ }\bibfield  {title} {\bibinfo {title} {Flow patterns around heart valves: A numerical method},\ }\href {https://doi.org/https://doi.org/10.1016/0021-9991(72)90065-4} {\bibfield  {journal} {\bibinfo  {journal} {Journal of Computational Physics}\ }\textbf {\bibinfo {volume} {10}},\ \bibinfo {pages} {252} (\bibinfo {year} {1972})}\BibitemShut {NoStop}%
\bibitem [{\citenamefont {Fadlun}\ \emph {et~al.}(2000)\citenamefont {Fadlun}, \citenamefont {Verzicco}, \citenamefont {Orlandi},\ and\ \citenamefont {Mohd-Yusof}}]{Fadlun2000}%
  \BibitemOpen
  \bibfield  {author} {\bibinfo {author} {\bibfnamefont {E.~A.}\ \bibnamefont {Fadlun}}, \bibinfo {author} {\bibfnamefont {R.}~\bibnamefont {Verzicco}}, \bibinfo {author} {\bibfnamefont {P.}~\bibnamefont {Orlandi}},\ and\ \bibinfo {author} {\bibfnamefont {J.}~\bibnamefont {Mohd-Yusof}},\ }\bibfield  {title} {\bibinfo {title} {Combined immersed-boundary finite-difference methods for three-dimensional complex flow simulations},\ }\href {https://doi.org/https://doi.org/10.1006/jcph.2000.6484} {\bibfield  {journal} {\bibinfo  {journal} {Journal of Computational Physics}\ }\textbf {\bibinfo {volume} {161}},\ \bibinfo {pages} {35} (\bibinfo {year} {2000})}\BibitemShut {NoStop}%
\bibitem [{\citenamefont {Orlandi}\ and\ \citenamefont {Leonardi}(2006)}]{Orlandi2006}%
  \BibitemOpen
  \bibfield  {author} {\bibinfo {author} {\bibfnamefont {P.}~\bibnamefont {Orlandi}}\ and\ \bibinfo {author} {\bibfnamefont {S.}~\bibnamefont {Leonardi}},\ }\bibfield  {title} {\bibinfo {title} {Dns of turbulent channel flows with two- and three-dimensional roughness},\ }\href {https://doi.org/10.1080/14685240600827526} {\bibfield  {journal} {\bibinfo  {journal} {Journal of Turbulence}\ }\textbf {\bibinfo {volume} {7}},\ \bibinfo {pages} {N73} (\bibinfo {year} {2006})}\BibitemShut {NoStop}%
\bibitem [{\citenamefont {Seo}\ and\ \citenamefont {Mittal}(2011)}]{Seo2011}%
  \BibitemOpen
  \bibfield  {author} {\bibinfo {author} {\bibfnamefont {J.~H.}\ \bibnamefont {Seo}}\ and\ \bibinfo {author} {\bibfnamefont {R.}~\bibnamefont {Mittal}},\ }\bibfield  {title} {\bibinfo {title} {A sharp-interface immersed boundary method with improved mass conservation and reduced spurious pressure oscillations},\ }\href {https://doi.org/https://doi.org/10.1016/j.jcp.2011.06.003} {\bibfield  {journal} {\bibinfo  {journal} {Journal of Computational Physics}\ }\textbf {\bibinfo {volume} {230}},\ \bibinfo {pages} {7347} (\bibinfo {year} {2011})}\BibitemShut {NoStop}%
\bibitem [{\citenamefont {Verzicco}(2023)}]{Verzicco2023}%
  \BibitemOpen
  \bibfield  {author} {\bibinfo {author} {\bibfnamefont {R.}~\bibnamefont {Verzicco}},\ }\bibfield  {title} {\bibinfo {title} {Immersed boundary methods: Historical perspective and future outlook},\ }\href {https://doi.org/10.1146/annurev-fluid-120720-022129} {\bibfield  {journal} {\bibinfo  {journal} {Annual Review of Fluid Mechanics}\ }\textbf {\bibinfo {volume} {55}},\ \bibinfo {pages} {129} (\bibinfo {year} {2023})}\BibitemShut {NoStop}%
\bibitem [{\citenamefont {Eaton}(2009)}]{Eaton2009}%
  \BibitemOpen
  \bibfield  {author} {\bibinfo {author} {\bibfnamefont {J.~K.}\ \bibnamefont {Eaton}},\ }\bibfield  {title} {\bibinfo {title} {Two-way coupled turbulence simulations of gas-particle flows using point-particle tracking},\ }\href {https://doi.org/10.1016/J.IJMULTIPHASEFLOW.2009.02.009} {\bibfield  {journal} {\bibinfo  {journal} {International Journal of Multiphase Flow}\ }\textbf {\bibinfo {volume} {35}},\ \bibinfo {pages} {792} (\bibinfo {year} {2009})}\BibitemShut {NoStop}%
\bibitem [{\citenamefont {Van~den Berg}\ \emph {et~al.}(2009)\citenamefont {Van~den Berg}, \citenamefont {Luther}, \citenamefont {Mazzitelli}, \citenamefont {Rensen}, \citenamefont {Toschi},\ and\ \citenamefont {Lohse}}]{VandenBerg2009}%
  \BibitemOpen
  \bibfield  {author} {\bibinfo {author} {\bibfnamefont {T.~H.}\ \bibnamefont {Van~den Berg}}, \bibinfo {author} {\bibfnamefont {S.}~\bibnamefont {Luther}}, \bibinfo {author} {\bibfnamefont {I.~M.}\ \bibnamefont {Mazzitelli}}, \bibinfo {author} {\bibfnamefont {J.~M.}\ \bibnamefont {Rensen}}, \bibinfo {author} {\bibfnamefont {F.}~\bibnamefont {Toschi}},\ and\ \bibinfo {author} {\bibfnamefont {D.}~\bibnamefont {Lohse}},\ }\bibfield  {title} {\bibinfo {title} {Turbulent bubbly flow},\ }\href {https://doi.org/10.1080/14685240500460782} {\bibfield  {journal} {\bibinfo  {journal} {Journal of Turbulence}\ }\textbf {\bibinfo {volume} {7}},\ \bibinfo {pages} {1} (\bibinfo {year} {2009})}\BibitemShut {NoStop}%
\bibitem [{\citenamefont {Monchaux}\ and\ \citenamefont {Dejoan}(2017)}]{Monchaux2017}%
  \BibitemOpen
  \bibfield  {author} {\bibinfo {author} {\bibfnamefont {R.}~\bibnamefont {Monchaux}}\ and\ \bibinfo {author} {\bibfnamefont {A.}~\bibnamefont {Dejoan}},\ }\bibfield  {title} {\bibinfo {title} {Settling velocity and preferential concentration of heavy particles under two-way coupling effects in homogeneous turbulence},\ }\href {https://doi.org/10.1103/PhysRevFluids.2.104302} {\bibfield  {journal} {\bibinfo  {journal} {Physical Review Fluids}\ }\textbf {\bibinfo {volume} {2}},\ \bibinfo {pages} {104302} (\bibinfo {year} {2017})}\BibitemShut {NoStop}%
\bibitem [{\citenamefont {Lamorgese}\ \emph {et~al.}(2005)\citenamefont {Lamorgese}, \citenamefont {Caughey},\ and\ \citenamefont {Pope}}]{Lamorgese2005}%
  \BibitemOpen
  \bibfield  {author} {\bibinfo {author} {\bibfnamefont {A.~G.}\ \bibnamefont {Lamorgese}}, \bibinfo {author} {\bibfnamefont {D.~A.}\ \bibnamefont {Caughey}},\ and\ \bibinfo {author} {\bibfnamefont {S.~B.}\ \bibnamefont {Pope}},\ }\bibfield  {title} {\bibinfo {title} {Direct numerical simulation of homogeneous turbulence with hyperviscosity},\ }\href {https://doi.org/10.1063/1.1833415} {\bibfield  {journal} {\bibinfo  {journal} {Physics of Fluids}\ }\textbf {\bibinfo {volume} {17}},\ \bibinfo {pages} {015106} (\bibinfo {year} {2005})}\BibitemShut {NoStop}%
\bibitem [{\citenamefont {Maxey}\ and\ \citenamefont {Riley}(1983)}]{Maxey1983}%
  \BibitemOpen
  \bibfield  {author} {\bibinfo {author} {\bibfnamefont {M.~R.}\ \bibnamefont {Maxey}}\ and\ \bibinfo {author} {\bibfnamefont {J.~J.}\ \bibnamefont {Riley}},\ }\bibfield  {title} {\bibinfo {title} {Equation of motion for a small rigid sphere in a nonuniform flow},\ }\href {https://doi.org/10.1063/1.864230} {\bibfield  {journal} {\bibinfo  {journal} {The Physics of Fluids}\ }\textbf {\bibinfo {volume} {26}},\ \bibinfo {pages} {883} (\bibinfo {year} {1983})}\BibitemShut {NoStop}%
\bibitem [{\citenamefont {van Hinsberg}\ \emph {et~al.}(2017)\citenamefont {van Hinsberg}, \citenamefont {Clercx},\ and\ \citenamefont {Toschi}}]{vanHinsberg2017}%
  \BibitemOpen
  \bibfield  {author} {\bibinfo {author} {\bibfnamefont {M.~A.~T.}\ \bibnamefont {van Hinsberg}}, \bibinfo {author} {\bibfnamefont {H.~J.~H.}\ \bibnamefont {Clercx}},\ and\ \bibinfo {author} {\bibfnamefont {F.}~\bibnamefont {Toschi}},\ }\bibfield  {title} {\bibinfo {title} {Enhanced settling of nonheavy inertial particles in homogeneous isotropic turbulence: The role of the pressure gradient and the basset history force},\ }\href {https://doi.org/10.1103/PhysRevE.95.023106} {\bibfield  {journal} {\bibinfo  {journal} {Physical Review E}\ }\textbf {\bibinfo {volume} {95}},\ \bibinfo {pages} {023106} (\bibinfo {year} {2017})}\BibitemShut {NoStop}%
\bibitem [{\citenamefont {van Hinsberg}\ \emph {et~al.}(2013)\citenamefont {van Hinsberg}, \citenamefont {ten Thije~Boonkkamp}, \citenamefont {Toschi},\ and\ \citenamefont {Clercx}}]{vanHinsberg2013}%
  \BibitemOpen
  \bibfield  {author} {\bibinfo {author} {\bibfnamefont {M.~A.~T.}\ \bibnamefont {van Hinsberg}}, \bibinfo {author} {\bibfnamefont {J.~H.~M.}\ \bibnamefont {ten Thije~Boonkkamp}}, \bibinfo {author} {\bibfnamefont {F.}~\bibnamefont {Toschi}},\ and\ \bibinfo {author} {\bibfnamefont {H.~J.~H.}\ \bibnamefont {Clercx}},\ }\bibfield  {title} {\bibinfo {title} {Optimal interpolation schemes for particle tracking in turbulence},\ }\href {https://doi.org/10.1103/PhysRevE.87.043307} {\bibfield  {journal} {\bibinfo  {journal} {Physical Review E}\ }\textbf {\bibinfo {volume} {87}},\ \bibinfo {pages} {043307} (\bibinfo {year} {2013})}\BibitemShut {NoStop}%
\bibitem [{Note1()}]{Note1}%
  \BibitemOpen
  \bibinfo {note} {Note that while physically, assuming Stokes drag on a sphere, the particle size $D/\eta $ becomes fixed as soon as one assumes a certain $\protect \textrm {St}$ and $\beta $, but numerically we can independently vary the particle size used in the two-way and four-way coupling to study its influence and to ensure that the particle remains sufficiently small.}\BibitemShut {Stop}%
\bibitem [{\citenamefont {Aliseda}\ and\ \citenamefont {Lasheras}(2011)}]{Aliseda2011}%
  \BibitemOpen
  \bibfield  {author} {\bibinfo {author} {\bibfnamefont {A.}~\bibnamefont {Aliseda}}\ and\ \bibinfo {author} {\bibfnamefont {J.~C.}\ \bibnamefont {Lasheras}},\ }\bibfield  {title} {\bibinfo {title} {Preferential concentration and rise velocity reduction of bubbles immersed in a homogeneous and isotropic turbulent flow},\ }\href {https://doi.org/10.1063/1.3626404} {\bibfield  {journal} {\bibinfo  {journal} {Physics of Fluids}\ }\textbf {\bibinfo {volume} {23}},\ \bibinfo {pages} {093301} (\bibinfo {year} {2011})}\BibitemShut {NoStop}%
\bibitem [{\citenamefont {Mercado}\ \emph {et~al.}(2012)\citenamefont {Mercado}, \citenamefont {Prakash}, \citenamefont {Tagawa}, \citenamefont {Sun}, \citenamefont {Lohse},\ and\ \citenamefont {for Turbulence~Research)}}]{Mercado2012}%
  \BibitemOpen
  \bibfield  {author} {\bibinfo {author} {\bibfnamefont {J.~M.}\ \bibnamefont {Mercado}}, \bibinfo {author} {\bibfnamefont {V.~N.}\ \bibnamefont {Prakash}}, \bibinfo {author} {\bibfnamefont {Y.}~\bibnamefont {Tagawa}}, \bibinfo {author} {\bibfnamefont {C.}~\bibnamefont {Sun}}, \bibinfo {author} {\bibfnamefont {D.}~\bibnamefont {Lohse}},\ and\ \bibinfo {author} {\bibfnamefont {I.~C.}\ \bibnamefont {for Turbulence~Research)}},\ }\bibfield  {title} {\bibinfo {title} {Lagrangian statistics of light particles in turbulence},\ }\href {https://doi.org/10.1063/1.4719148} {\bibfield  {journal} {\bibinfo  {journal} {Physics of Fluids}\ }\textbf {\bibinfo {volume} {24}},\ \bibinfo {pages} {055106} (\bibinfo {year} {2012})}\BibitemShut {NoStop}%
\bibitem [{\citenamefont {Chouippe}\ \emph {et~al.}(2014)\citenamefont {Chouippe}, \citenamefont {Climent}, \citenamefont {Legendre},\ and\ \citenamefont {Gabillet}}]{Chouippe2014}%
  \BibitemOpen
  \bibfield  {author} {\bibinfo {author} {\bibfnamefont {A.}~\bibnamefont {Chouippe}}, \bibinfo {author} {\bibfnamefont {E.}~\bibnamefont {Climent}}, \bibinfo {author} {\bibfnamefont {D.}~\bibnamefont {Legendre}},\ and\ \bibinfo {author} {\bibfnamefont {C.}~\bibnamefont {Gabillet}},\ }\bibfield  {title} {\bibinfo {title} {Numerical simulation of bubble dispersion in turbulent taylor-couette flow},\ }\href {https://doi.org/10.1063/1.4871728} {\bibfield  {journal} {\bibinfo  {journal} {Physics of Fluids}\ }\textbf {\bibinfo {volume} {26}},\ \bibinfo {pages} {043304} (\bibinfo {year} {2014})}\BibitemShut {NoStop}%
\bibitem [{\citenamefont {Mathai}\ \emph {et~al.}(2016)\citenamefont {Mathai}, \citenamefont {Calzavarini}, \citenamefont {Brons}, \citenamefont {Sun},\ and\ \citenamefont {Lohse}}]{Mathai2016}%
  \BibitemOpen
  \bibfield  {author} {\bibinfo {author} {\bibfnamefont {V.}~\bibnamefont {Mathai}}, \bibinfo {author} {\bibfnamefont {E.}~\bibnamefont {Calzavarini}}, \bibinfo {author} {\bibfnamefont {J.}~\bibnamefont {Brons}}, \bibinfo {author} {\bibfnamefont {C.}~\bibnamefont {Sun}},\ and\ \bibinfo {author} {\bibfnamefont {D.}~\bibnamefont {Lohse}},\ }\bibfield  {title} {\bibinfo {title} {Microbubbles and microparticles are not faithful tracers of turbulent acceleration},\ }\href {https://doi.org/10.1103/PhysRevLett.117.024501} {\bibfield  {journal} {\bibinfo  {journal} {Physical Review Letters}\ }\textbf {\bibinfo {volume} {117}},\ \bibinfo {pages} {024501} (\bibinfo {year} {2016})}\BibitemShut {NoStop}%
\bibitem [{\citenamefont {Loisy}\ and\ \citenamefont {Naso}(2017)}]{Loisy2017}%
  \BibitemOpen
  \bibfield  {author} {\bibinfo {author} {\bibfnamefont {A.}~\bibnamefont {Loisy}}\ and\ \bibinfo {author} {\bibfnamefont {A.}~\bibnamefont {Naso}},\ }\bibfield  {title} {\bibinfo {title} {Interaction between a large buoyant bubble and turbulence},\ }\href {https://doi.org/10.1103/PhysRevFluids.2.014606} {\bibfield  {journal} {\bibinfo  {journal} {Physical Review Fluids}\ }\textbf {\bibinfo {volume} {2}},\ \bibinfo {pages} {014606} (\bibinfo {year} {2017})}\BibitemShut {NoStop}%
\bibitem [{\citenamefont {Mathai}\ \emph {et~al.}(2018)\citenamefont {Mathai}, \citenamefont {Huisman}, \citenamefont {Sun}, \citenamefont {Lohse},\ and\ \citenamefont {Bourgoin}}]{Mathai2018}%
  \BibitemOpen
  \bibfield  {author} {\bibinfo {author} {\bibfnamefont {V.}~\bibnamefont {Mathai}}, \bibinfo {author} {\bibfnamefont {S.~G.}\ \bibnamefont {Huisman}}, \bibinfo {author} {\bibfnamefont {C.}~\bibnamefont {Sun}}, \bibinfo {author} {\bibfnamefont {D.}~\bibnamefont {Lohse}},\ and\ \bibinfo {author} {\bibfnamefont {M.}~\bibnamefont {Bourgoin}},\ }\bibfield  {title} {\bibinfo {title} {Dispersion of air bubbles in isotropic turbulence},\ }\href {https://doi.org/10.1103/PhysRevLett.121.054501} {\bibfield  {journal} {\bibinfo  {journal} {Physical Review Letters}\ }\textbf {\bibinfo {volume} {121}},\ \bibinfo {pages} {054501} (\bibinfo {year} {2018})}\BibitemShut {NoStop}%
\bibitem [{\citenamefont {Boivin}\ \emph {et~al.}(1998)\citenamefont {Boivin}, \citenamefont {Simonin},\ and\ \citenamefont {Squires}}]{Boivin1998}%
  \BibitemOpen
  \bibfield  {author} {\bibinfo {author} {\bibfnamefont {M.}~\bibnamefont {Boivin}}, \bibinfo {author} {\bibfnamefont {O.}~\bibnamefont {Simonin}},\ and\ \bibinfo {author} {\bibfnamefont {K.~D.}\ \bibnamefont {Squires}},\ }\bibfield  {title} {\bibinfo {title} {Direct numerical simulation of turbulence modulation by particles in isotropic turbulence},\ }\href {https://doi.org/10.1017/S0022112098002821} {\bibfield  {journal} {\bibinfo  {journal} {Journal of Fluid Mechanics}\ }\textbf {\bibinfo {volume} {375}},\ \bibinfo {pages} {235} (\bibinfo {year} {1998})}\BibitemShut {NoStop}%
\bibitem [{\citenamefont {Mazzitelli}\ \emph {et~al.}(2003)\citenamefont {Mazzitelli}, \citenamefont {Lohse},\ and\ \citenamefont {Toschi}}]{Mazzitelli2003}%
  \BibitemOpen
  \bibfield  {author} {\bibinfo {author} {\bibfnamefont {I.~M.}\ \bibnamefont {Mazzitelli}}, \bibinfo {author} {\bibfnamefont {D.}~\bibnamefont {Lohse}},\ and\ \bibinfo {author} {\bibfnamefont {F.}~\bibnamefont {Toschi}},\ }\bibfield  {title} {\bibinfo {title} {On the relevance of the lift force in bubbly turbulence},\ }\href {https://doi.org/10.1017/S0022112003004877} {\bibfield  {journal} {\bibinfo  {journal} {Journal of Fluid Mechanics}\ }\textbf {\bibinfo {volume} {488}},\ \bibinfo {pages} {283–313} (\bibinfo {year} {2003})}\BibitemShut {NoStop}%
\bibitem [{\citenamefont {Maxey}\ \emph {et~al.}(1994)\citenamefont {Maxey}, \citenamefont {Chang},\ and\ \citenamefont {Wang}}]{Maxey1994}%
  \BibitemOpen
  \bibfield  {author} {\bibinfo {author} {\bibfnamefont {M.~R.}\ \bibnamefont {Maxey}}, \bibinfo {author} {\bibfnamefont {E.~J.}\ \bibnamefont {Chang}},\ and\ \bibinfo {author} {\bibfnamefont {L.~P.}\ \bibnamefont {Wang}},\ }\bibfield  {title} {\bibinfo {title} {Simulation of interactions between microbubbles and turbulent flows},\ }\href {https://doi.org/10.1115/1.3124443} {\bibfield  {journal} {\bibinfo  {journal} {Applied Mechanics Reviews}\ }\textbf {\bibinfo {volume} {47}},\ \bibinfo {pages} {S70} (\bibinfo {year} {1994})}\BibitemShut {NoStop}%
\bibitem [{\citenamefont {Tom}\ \emph {et~al.}(2022)\citenamefont {Tom}, \citenamefont {Carbone},\ and\ \citenamefont {Bragg}}]{Tom2022}%
  \BibitemOpen
  \bibfield  {author} {\bibinfo {author} {\bibfnamefont {J.}~\bibnamefont {Tom}}, \bibinfo {author} {\bibfnamefont {M.}~\bibnamefont {Carbone}},\ and\ \bibinfo {author} {\bibfnamefont {A.~D.}\ \bibnamefont {Bragg}},\ }\bibfield  {title} {\bibinfo {title} {How does two-way coupling modify particle settling and the role of multiscale preferential sweeping?},\ }\href {https://doi.org/10.1017/jfm.2022.615} {\bibfield  {journal} {\bibinfo  {journal} {Journal of Fluid Mechanics}\ }\textbf {\bibinfo {volume} {947}},\ \bibinfo {pages} {A7} (\bibinfo {year} {2022})}\BibitemShut {NoStop}%
\bibitem [{\citenamefont {Pekurovsky}(2012)}]{Pekurovsky2012}%
  \BibitemOpen
  \bibfield  {author} {\bibinfo {author} {\bibfnamefont {D.}~\bibnamefont {Pekurovsky}},\ }\bibfield  {title} {\bibinfo {title} {{P3DFFT}: A framework for parallel computations of fourier transforms in three dimensions},\ }\href {https://doi.org/10.1137/11082748X} {\bibfield  {journal} {\bibinfo  {journal} {SIAM Journal on Scientific Computing}\ }\textbf {\bibinfo {volume} {34}},\ \bibinfo {pages} {C192} (\bibinfo {year} {2012})}\BibitemShut {NoStop}%
\bibitem [{\citenamefont {Allen}\ and\ \citenamefont {Tildesley}(1987)}]{Allen1987}%
  \BibitemOpen
  \bibfield  {author} {\bibinfo {author} {\bibfnamefont {M.~P.}\ \bibnamefont {Allen}}\ and\ \bibinfo {author} {\bibfnamefont {D.~J.}\ \bibnamefont {Tildesley}},\ }\href@noop {} {\emph {\bibinfo {title} {Computer simulation of liquids}}}\ (\bibinfo  {publisher} {Oxford University Press},\ \bibinfo {year} {1987})\BibitemShut {NoStop}%
\bibitem [{\citenamefont {Lance}\ and\ \citenamefont {Bataille}(1991)}]{Lance1991}%
  \BibitemOpen
  \bibfield  {author} {\bibinfo {author} {\bibfnamefont {M.}~\bibnamefont {Lance}}\ and\ \bibinfo {author} {\bibfnamefont {J.}~\bibnamefont {Bataille}},\ }\bibfield  {title} {\bibinfo {title} {Turbulence in the liquid phase of a uniform bubbly air–water flow},\ }\href {https://doi.org/10.1017/S0022112091001015} {\bibfield  {journal} {\bibinfo  {journal} {Journal of Fluid Mechanics}\ }\textbf {\bibinfo {volume} {222}},\ \bibinfo {pages} {95–118} (\bibinfo {year} {1991})}\BibitemShut {NoStop}%
\bibitem [{\citenamefont {Riboux}\ \emph {et~al.}(2010)\citenamefont {Riboux}, \citenamefont {Risso},\ and\ \citenamefont {Legendre}}]{Riboux2010}%
  \BibitemOpen
  \bibfield  {author} {\bibinfo {author} {\bibfnamefont {G.}~\bibnamefont {Riboux}}, \bibinfo {author} {\bibfnamefont {F.}~\bibnamefont {Risso}},\ and\ \bibinfo {author} {\bibfnamefont {D.}~\bibnamefont {Legendre}},\ }\bibfield  {title} {\bibinfo {title} {Experimental characterization of the agitation generated by bubbles rising at high reynolds number},\ }\href {https://doi.org/10.1017/S0022112009992084} {\bibfield  {journal} {\bibinfo  {journal} {Journal of Fluid Mechanics}\ }\textbf {\bibinfo {volume} {643}},\ \bibinfo {pages} {509–539} (\bibinfo {year} {2010})}\BibitemShut {NoStop}%
\bibitem [{\citenamefont {Alméras}\ \emph {et~al.}(2017)\citenamefont {Alméras}, \citenamefont {Mathai}, \citenamefont {Lohse},\ and\ \citenamefont {Sun}}]{Almeras2017}%
  \BibitemOpen
  \bibfield  {author} {\bibinfo {author} {\bibfnamefont {E.}~\bibnamefont {Alméras}}, \bibinfo {author} {\bibfnamefont {V.}~\bibnamefont {Mathai}}, \bibinfo {author} {\bibfnamefont {D.}~\bibnamefont {Lohse}},\ and\ \bibinfo {author} {\bibfnamefont {C.}~\bibnamefont {Sun}},\ }\bibfield  {title} {\bibinfo {title} {Experimental investigation of the turbulence induced by a bubble swarm rising within incident turbulence},\ }\href {https://doi.org/10.1017/jfm.2017.410} {\bibfield  {journal} {\bibinfo  {journal} {Journal of Fluid Mechanics}\ }\textbf {\bibinfo {volume} {825}},\ \bibinfo {pages} {1091–1112} (\bibinfo {year} {2017})}\BibitemShut {NoStop}%
\bibitem [{\citenamefont {Andersson}\ \emph {et~al.}(2012)\citenamefont {Andersson}, \citenamefont {Zhao},\ and\ \citenamefont {Barri}}]{Andersson2012}%
  \BibitemOpen
  \bibfield  {author} {\bibinfo {author} {\bibfnamefont {H.~I.}\ \bibnamefont {Andersson}}, \bibinfo {author} {\bibfnamefont {L.}~\bibnamefont {Zhao}},\ and\ \bibinfo {author} {\bibfnamefont {M.}~\bibnamefont {Barri}},\ }\bibfield  {title} {\bibinfo {title} {Torque-coupling and particle–turbulence interactions},\ }\href {https://doi.org/10.1017/JFM.2012.44} {\bibfield  {journal} {\bibinfo  {journal} {Journal of Fluid Mechanics}\ }\textbf {\bibinfo {volume} {696}},\ \bibinfo {pages} {319} (\bibinfo {year} {2012})}\BibitemShut {NoStop}%
\bibitem [{\citenamefont {Frigo}\ and\ \citenamefont {Johnson}(2005)}]{FFTW05}%
  \BibitemOpen
  \bibfield  {author} {\bibinfo {author} {\bibfnamefont {M.}~\bibnamefont {Frigo}}\ and\ \bibinfo {author} {\bibfnamefont {S.}~\bibnamefont {Johnson}},\ }\bibfield  {title} {\bibinfo {title} {The design and implementation of {FFTW3}},\ }\href {https://doi.org/10.1109/JPROC.2004.840301} {\bibfield  {journal} {\bibinfo  {journal} {Proceedings of the IEEE}\ }\textbf {\bibinfo {volume} {93}},\ \bibinfo {pages} {216} (\bibinfo {year} {2005})}\BibitemShut {NoStop}%
\end{thebibliography}


\end{document}